\documentclass[twocolumn,twocolappendix]{aastex631}
\pdfoutput=1 
\usepackage{amsmath,amstext}
\usepackage[T1]{fontenc}
\usepackage{ae,aecompl}
\usepackage[utf8]{inputenc}
\usepackage{apjfonts} 
\usepackage[figure,figure*]{hypcap}
\usepackage{enumitem}
\usepackage[hang]{footmisc}
\input{hyperlink-year-only-natbib-patch}
\allowdisplaybreaks
\graphicspath{{updated_figures_new/}{updated_figures/}{Figures/}}

\newcommand{\pc}{\ensuremath{{\rm pc}}}
\newcommand{\kpc}{\ensuremath{{\rm kpc}}}
\newcommand{\mpc}{\ensuremath{{\rm Mpc}}}
\newcommand{\kms}{\ensuremath{{\rm km\,s^{-1}}}}

\newcommand{\Myr}{\ensuremath{{\rm Myr}}}
\newcommand{\Gyr}{\ensuremath{{\rm Gyr}}}

\newcommand{\msun}{\ensuremath{{M_{\odot}}}}

\newcommand{\rvir}{\ensuremath{{R_{\rm vir}}}}

\shorttitle{Milky Way-est Simulations}
\shortauthors{Buch, Nadler, Wechsler, \& Mao}

\begin{document}

\title{Milky Way-est: Cosmological Zoom-in Simulations with Large Magellanic Cloud and Gaia--Sausage--Enceladus Analogs}

\author[0000-0002-6635-4712]{Deveshi~Buch}
\affiliation{Kavli Institute for Particle Astrophysics \& Cosmology, P. O. Box 2450, Stanford University, Stanford, CA 94305, USA}
\affiliation{Department of Computer Science, Stanford University, 353 Jane Stanford Way, Stanford, CA 94305, USA}
\author[0000-0002-1182-3825]{Ethan~O.~Nadler}
\affiliation{Carnegie Observatories, 813 Santa Barbara Street, Pasadena, CA 91101, USA}
\affiliation{Department of Physics $\&$ Astronomy, University of Southern California, Los Angeles, CA 90007, USA}
\author[0000-0003-2229-011X]{Risa~H.~Wechsler}
\affiliation{Kavli Institute for Particle Astrophysics \& Cosmology, P. O. Box 2450, Stanford University, Stanford, CA 94305, USA}
\affiliation{Department of Physics, Stanford University, 382 Via Pueblo Mall, Stanford, CA 94305, USA}
\affiliation{SLAC National Accelerator Laboratory, Menlo Park, CA 94025, USA}
\author[0000-0002-1200-0820]{Yao-Yuan~Mao}
\affiliation{Department of Physics and Astronomy, University of Utah, Salt Lake City, UT 84112, USA}

\correspondingauthor{Deveshi Buch}
\email{deveshi@cs.stanford.edu}

\begin{abstract}
We present Milky Way-est, a suite of 20 cosmological cold-dark-matter-only zoom-in simulations of Milky Way (MW)-like host halos. Milky Way-est hosts are selected such that they (i) are consistent with the MW's measured halo mass and concentration, (ii) accrete a Large Magellanic Cloud (LMC)-like ($\approx 10^{11}~\msun$) subhalo within the last $2~\Gyr$ on a realistic orbit, placing them near $50~\kpc$ from the host center at $z\approx 0$, and (iii) undergo a $>$1:5 sub-to-host halo mass ratio merger with a Gaia--Sausage--Enceladus (GSE)-like system at early times ($0.67<z<3$). Hosts satisfying these LMC and GSE constraints constitute $< 1\%$ of all halos in the MW's mass range, and their total masses grow rapidly at late times due to LMC analog accretion. Compared to hosts of a similar final halo mass that are not selected to include LMC and GSE analogs, Milky Way-est hosts contain $22\%$ more subhalos with present-day virial masses above $10^8~\msun$ throughout the virial radius, on average. This enhancement reaches $\approx 80\%$ in the inner $100~\mathrm{kpc}$ and is largely, if not entirely, due to LMC-associated subhalos. These systems also induce spatial anisotropy in Milky Way-est subhalo populations, with $\approx 60\%$ of the total subhalo population within $100~\kpc$ found in the current direction of the LMC. Meanwhile, we find that GSE-associated subhalos do not significantly contribute to present-day Milky Way-est subhalo populations. These results provide context for our Galaxy's dark matter structure and subhalo population and will help interpret a range of measurements that are currently only possible in the MW.
\end{abstract}

\keywords{\href{http://astrothesaurus.org/uat/353}{Dark matter (353)}; \href{http://astrothesaurus.org/uat/903}{Large Magellanic Cloud (903)}; \href{http://astrothesaurus.org/uat/1049}{Milky Way dark matter halo (1049)}; \href{http://astrothesaurus.org/uat/1083}{N-body simulations (1083)}}


\section{Introduction}
\label{Introduction}

The Milky Way (MW)'s dark matter distribution is a cornerstone of near-field cosmology. In recent years, astrometric measurements \citep{Gardner2021103904}, photometric surveys \citep{Drlica-Wagner191203302,Drlica-Wagner220316565}, and spectroscopic campaigns \citep{APOGSEE2017,Naidu200608625,Cooper220808514} have qualitatively advanced our understanding of the events that contributed to the buildup of dark matter in the MW. These observations reveal that the MW’s most massive surviving and disrupted satellites---including the recently accreted Large Magellanic Cloud (LMC) system (\citealt{vanderMarel0205161,besla2007,Kallivayalil13010832,Vasiliev230409136}) and the ancient Gaia--Sausage--Enceladus (GSE) merger (\citealt{Belokurov180203414,Helmi180606038})---play a crucial role in shaping its dark matter distribution and substructure (e.g., \citealt{Helmi_streams,ConroyNature}).

In particular, the recently accreted LMC drives the MW's dark matter distribution away from dynamical equilibrium, inducing wakes, deformation, and reflex motion that have recently been detected (e.g., \citealt{Garavito-Camargo190205089,Garavity-Camargo201000816,Petersen201110581,ConroyNature,Lilleengen220501688}). The LMC also brings its own substructure into the MW, reflected by the existence of kinematically associated satellite galaxies (e.g., \citealt{Kallivayalil_2018,Patel200101746,Samuel201008571}) and spatial anisotropy in the distribution of MW satellite galaxies (e.g., \citealt{Jethwa160304420, Nadler191203303}). Meanwhile, the ancient GSE merger is a key event in the formation of the MW galaxy, as evidenced by the chemo-kinematic structure of the MW's disk (e.g., \citealt{Grand200106009,Belokurov220304980,Semenov230609398,Dillamore230908658}). GSE also significantly impacts the dynamical structure of the MW stellar halo, which is dominated by stars on radial orbits deposited by this merger at early times (e.g., \citealt{Fattahi181007779,Dillamore210913244,Han220804327}). Finally, the Sagittarius (Sgr) dwarf galaxy (\citealt{Ibata1994}) fell into the MW between the GSE and LMC and is tidally disrupting today (e.g., \citealt{Vasiliev200602929}). Sgr plays an important role in shaping the MW's stellar halo, stellar streams population, and disk, and may have had an infall mass comparable to the LMC (e.g., \citealt{Kruijssen200301119}).

From a theoretical perspective, cosmological simulations provide detailed predictions for the growth and structure of MW-mass halos, including their subhalo populations (e.g., \citealt{SpringelAquarius, Diemand08051244, MWW2015, GriffenCaterpillar, SawalaApostle, WetzelLatte, Poole-McKenzieARTEMIS,Nadler_2023}). Most studies that use simulations to interpret MW observations select hosts with masses and concentrations consistent with those inferred for the MW halo, regardless of merger history. In hydrodynamic contexts, selection often includes central galaxy properties (e.g., stellar mass and morphology). Other simulations require MW analogs to reside in environments similar to the Local Group or Local Volume (e.g., \citealt{Garrison-Kimmel13106746, Carlesi160203919, hestiapaper, Sawala210312073}).

Motivated by the observational advances discussed above, recent simulations have focused on halos that satisfy additional constraints on the MW's merger history. Considering the LMC, several authors have analyzed small numbers of MW-mass hosts with LMC analog subhalos using gravity-only (e.g., \citealt{Erkal190709484, Sales160503574}) or hydrodynamic simulations (e.g., \citealt{Barry230305527,Smith-Orlik230204281,Arora230915998}). Considering GSE and other early accretion events, \cite{Bignone190807080} and \cite{Bose190904039} study hosts in cosmological hydrodynamic simulations that include GSE analogs, \cite{Rey221115689} present zoom-in simulations that are engineered to include GSE analogs with controlled properties, and \cite{Bethencourt231011300} present a high-resolution hydrodynamical simulation that includes analogs of the GSE and the early Kraken and Sequoia mergers. Beyond individual constraints, \cite{Evans200504969} analyze MW-like hosts with both LMC and GSE analogs using large-volume cosmological simulations, and \cite{Nadler191203303} study zoom ins of two MW-mass hosts with analogs of both systems. 

Despite this progress, no existing simulation suite is designed to study MW-mass hosts that accrete an LMC analog at late times and have an early accretion history consistent with that of the MW (e.g., by undergoing a major merger with a GSE analog at early times). Thus, it is timely to develop high-resolution cosmological simulations that incorporate such constraints at both early and late times.

Here, we introduce Milky Way-est, a suite of $20$ cosmological cold-dark-matter-only zoom-in simulations focused on hosts that are consistent with the MW's halo mass and concentration measurements and that accrete both LMC and GSE analogs. While we do not explicitly include an Sgr constraint in order to obtain a reasonably large host sample, we note that nearly half of our hosts accrete a massive subhalo between GSE disruption and LMC infall that merges by $z=0$. To benchmark our results, we compare Milky Way-est to the Symphony Milky Way-mass zoom-in simulation suite (hereafter Symphony MW; \citealt{MWW2015,Nadler_2023}). Symphony MW includes 45 MW-mass hosts, run with the same resolution and numerical settings as Milky Way-est, which are not selected to include LMC and GSE analogs. Comparing Milky Way-est and Symphony MW hosts can, therefore, help isolate the effects of our GSE and LMC constraints on hosts' mass accretion histories (MAHs) and subhalo populations. We will show that Milky Way-est hosts' formation times, subhalo mass functions (SHMFs), and subhalo radial and spatial distributions significantly differ from Symphony MW. Thus, it is important to model both the LMC and GSE to accurately interpret measurements of the MW's dark matter distribution and substructure.

Milky Way-est simulations resolve subhalos down to present-day masses of $\approx 10^8~\msun$, near the threshold of galaxy formation \citep{Benitez-Llambay200406124,Nadler191203303}; detailed comparisons between Milky Way-est and Symphony MW will therefore help contextualize upcoming observations of the MW satellite population, which probe galaxy formation and dark matter physics \citep{Nadler240110318}. 
More generally, Milky Way-est provides a realistic setting for testing empirical galaxy--halo connection models (e.g., \citealt{Nadler180905542,Wang210211876}), conducting zoom-in simulations of galaxy formation (e.g., \citealt{WetzelLatte,Font200401914,Applebaum200811207}), and exploring physics beyond the cold-dark-matter paradigm (e.g., \citealt{Nadler200800022,Nadler210912120,Mau220111740,Yang221113768}) in a setting that facilitates accurate predictions for near-field cosmology.

This paper is organized as follows. Section~\ref{sec:sims} describes our simulations, including host halo selection criteria and the Milky Way-est sample. In Section~\ref{sec:characteristics}, we present the properties of Milky Way-est hosts, LMC analogs, and GSE analogs. In Section~\ref{sec:results}, we study Milky Way-est host halo MAHs and subhalo populations compared to Symphony MW. We discuss the implications of our results, areas for future work, and caveats in Section~\ref{sec:future}. We conclude in Section~\ref{sec:conclusion}. We list the names and properties of Milky Way-est systems (including their LMC and GSE analog halos) in Appendix~\ref{sec:host_properties}, and a convergence test using one higher-resolution Milky Way-est resimulation is presented in Appendix~\ref{sec:convergence}.

\begin{figure*}[htb!]
\hspace{-5mm}
\includegraphics[trim={0 1.9cm 0 0},width=\textwidth]{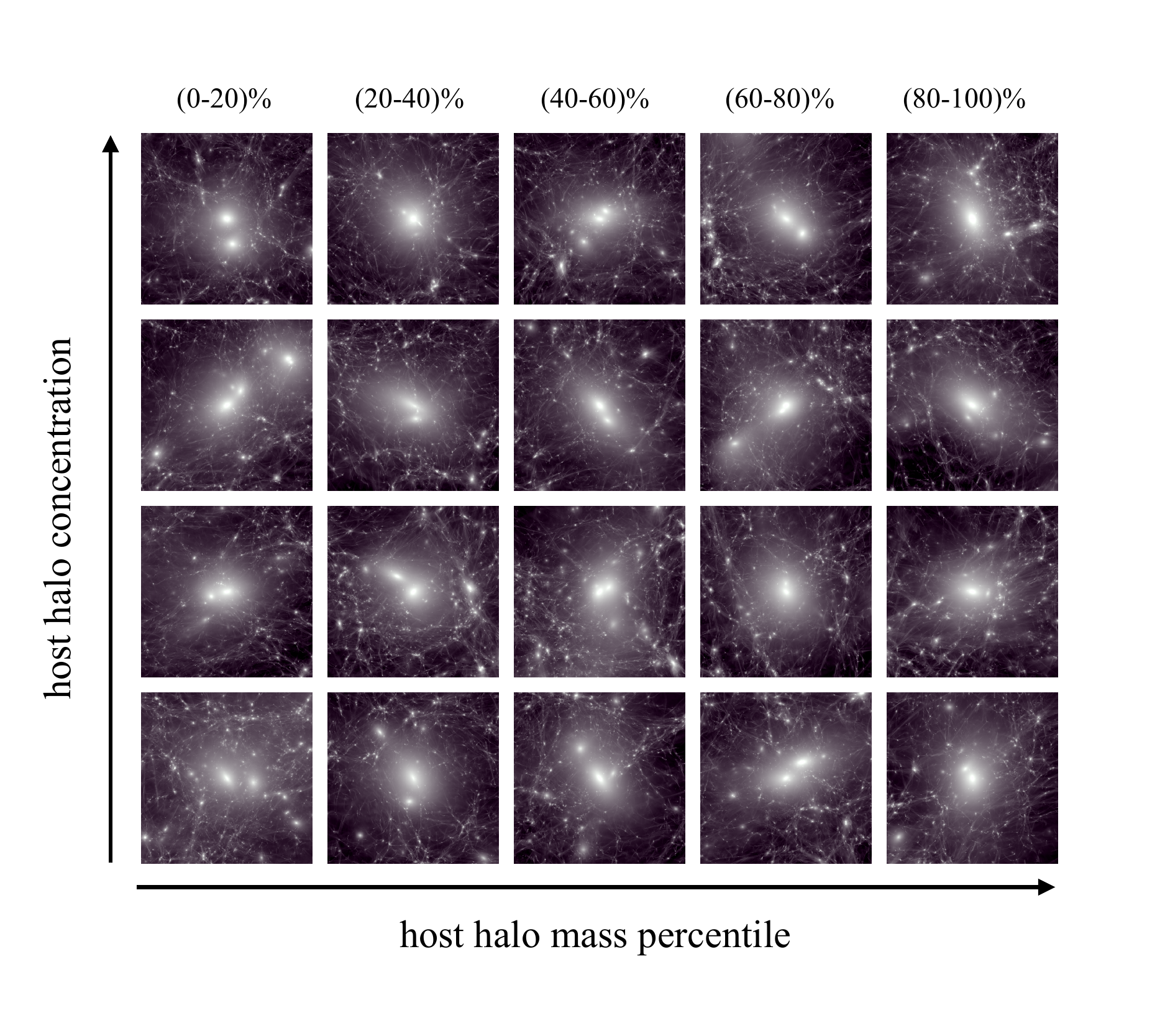}
\caption{Projected dark matter density of Milky Way-est hosts. Systems are arranged horizontally by host mass percentile and vertically by host concentration within each mass bin. Visualizations are created using the phase-space tessellation method \citep{KAEHLER201768,Kaehler:2018:2470-1173:333}. Each simulation is visualized at the snapshot when its LMC analog is nearest to $50~\kpc$ from the host center (see Section~\ref{sec:sample}).}
\label{fig:mwest_grid}
\end{figure*}


\section{Simulations}
\label{sec:sims}

We begin by describing our simulations, including technical details (Section~\ref{sec:technical}), host halo selection criteria (Section~\ref{sec:criteria}), the Milky Way-est sample that we resimulate using the zoom-in technique (Section~\ref{sec:sample}), and the MW-mass sample, Symphony MW, that we use in comparative analyses with our suite (Section~\ref{sec:symphonysample}).

\subsection{Simulation Setup}
\label{sec:technical}

Following \cite{Nadler_2023}, all of our simulations use $h = 0.7$, $\Omega_{\rm m} = 0.286$, $\Omega_{\Lambda} = 0.714$, $\sigma_8 = 0.82$, and $n_s=0.96$ (similar to WMAP 9; \citealt{Hinshaw_2013}). Halo catalogs and merger trees are generated using {\sc Rockstar} and {\sc Consistent-Trees} \citep{Behroozi11104372,Behroozi11104370}. We define ``peak'' as the highest value of a quantity along a halo's main branch, and we measure concentration as $c_{\mathrm{vir}}\equiv R_{\mathrm{vir}}/R_s$, where $R_s$ is the Navarro--Frenk--White (\citealt{Navarro9611107}) scale radius fit by {\sc Rockstar}.

Milky Way-est hosts are selected from Chinchilla \textsc{c125-1024}, a cosmological dark-matter-only simulation run with a box length of $125~{\rm Mpc}~h^{-1}$ and $1024^3$ particles (see \citealt{MWW2015}; \citealt{Wang210211876}, for details). We select hosts based on several criteria, including host halo properties that are consistent with inferred properties of the MW's dark matter halo and the existence of realistic LMC and GSE analogs, as described in Section~\ref{sec:criteria}. We generate zoom-in initial conditions with four refinement regions, yielding an equivalent of $8192$ particles per side in the most refined region using {\sc MUSIC} \citep{Hahn1103.6031}. The highest-resolution region corresponds to the Lagrangian volume of particles within 10 times the $z=0$ virial radius of each host ($R_{\mathrm{vir,host}}$) in \textsc{c125-1024} (with the exception of Halo756, for which we use $7R_{\mathrm{vir,host}}$ to avoid an expensive resimulation).\footnote{Virial quantities are calculated using the \cite{Bryan_Norman} overdensity, which corresponds to $\Delta_{\mathrm{vir}}\approx 99$ times the critical density of the universe in our cosmology at $z=0$.} 

We run zoom-in simulations using {\sc GADGET-2} \citep{springel2005gadget2}, with numerical parameters matched to those for Symphony MW (see \citealt{MWW2015,Nadler_2023} for details). The dark matter particle mass in the highest-resolution region is $m_{\mathrm{part}}=4.0\times 10^5~\msun$ and the comoving Plummer-equivalent gravitational softening is $\epsilon=170~\pc~h^{-1}$. We also verify convergence using a resimulation of one Milky Way-est host with $8\times$~smaller particle mass and $2\times$~smaller softening (see Appendix~\ref{sec:convergence}). We initialize each zoom in at $z=99$ and save 236 snapshots until $z=0$. At late times, the typical time spacing between snapshots is $25~\Myr$; as a result, we run some Milky Way-est zoom ins slightly past $z=0$ to ensure that our LMC analogs reach a distance of $\approx 50~\kpc$ (see Section~\ref{sec:sample} for details). We assume the same convergence properties derived in \cite{Nadler_2023} and apply a cut on a present-day subhalo virial mass of $M_{\mathrm{sub}}>300\times m_{\mathrm{part}}=1.2\times 10^8~ \msun$ when analyzing individual simulations. We conservatively cut on $M_{\mathrm{sub}}/M_{\mathrm{host}}>1.4\times 10^{-4}$ when analyzing the entire suite, where $M_{\mathrm{host}}$ is defined as the total dark matter mass within $R_{\mathrm{vir,host}}$, including all subhalos. This threshold corresponds to $300\times m_{\mathrm{part}}$ for the lowest-mass Milky Way-est host.

\subsection{Milky Way-est Host Halo Selection} 
\label{sec:criteria}

We select hosts in \textsc{c125-1024} that satisfy all of the following criteria:
\begin{enumerate}[label=(\Roman*),topsep=0pt,itemsep=-1ex,partopsep=1ex,parsep=1ex]
    \item Host mass, $M_{\mathrm{host}}$, between $1\times 10^{12}$ and $1.8\times10^{12} ~\msun$ (corresponding to the $2\sigma$ range of MW host halo mass derived in \citealt{Callingham180810456}).
    \item Host concentration, $c_{\mathrm{host}}$, between 7 and 16 (corresponding to the $2\sigma$ range of MW host halo concentration derived in \citealt{Callingham180810456}).
    \item An LMC analog subhalo that satisfies the following:
    \begin{enumerate}[label=(\alph*),topsep=-1pt,itemsep=-1ex,partopsep=1ex,parsep=1ex]
        \item Maximum circular velocity $V_{\text{max,sub}} > 55~\kms$ (\citealt{vanderMarel13054641}; if multiple exist, we choose the subhalo with the highest peak maximum circular velocity, $V_{\text{peak,sub}}$).
        \item Scale factor of most recent infall into the MW virial radius $a_{\mathrm{infall}}>0.86$ ($z_{\mathrm{infall}}<0.16$, corresponding to a lookback time within $2~\Gyr$ of $z=0$; \citealt{Kallivayalil13010832}).
        \item Distance to host at $z=0$, $r_{\mathrm{LMC}}$, between 30 and 70 kpc (e.g., \citealt{Pietrzynski2019}).
    \end{enumerate}
    \item A GSE analog subhalo that merges with the MW host between $0.25 < a_{\mathrm{disrupt}} < 0.6$ (i.e., $0.67<z_{\mathrm{disrupt}}<3$ or between $6$ and $11.5~\Gyr$ ago) with $M_{\mathrm{sub}}/M_{\mathrm{host}}>0.2$ when the GSE analog achieves its peak mass (e.g., \citealt{Helmi_streams,Naidu210303251}).
\end{enumerate}
The selection criteria are designed to bracket observationally inferred properties of the MW, LMC, and GSE halos. Given these constraints, we find that $< 1\%$ of MW-mass halos undergo both accretion events, adding to previous findings regarding the rarity of the LMC/Small Magellanic Cloud (SMC; e.g., \citealt{BoylanKolchin09114484,Busha10116373,Liu10112255}).

Starting with the host, our cuts encompass most MW halo mass and concentration estimates in the literature (e.g., see \citealt{Wang191202599}, for a review). Note that we measure halo mass and concentration in the absence of baryons because we work with gravity-only simulations. Including baryons may affect the values of both quantities (e.g., \citealt{Cautun191104557}), and it would be interesting to explore this effect with hydrodynamic resimulations of Milky Way-est hosts. Regardless, the MW halo properties from \cite{Callingham180810456} that we use account for this effect.

\begin{deluxetable}{l c c}[t!]
\centering
\tabletypesize{\footnotesize}
\tablecaption{\label{tab:1} {Percentage of MW-mass halos from \textsc{c125-1024} that pass each Milky Way-est selection cut, corresponding to the criteria from Section~\ref{sec:criteria}. For each criterion, the percentage of halos that pass only the previous cut and the remaining percentage of MW-mass halos are shown.}}
\tablehead{\colhead{Criterion} & \colhead{\% of Previous} & \colhead{\% of MW Mass}}
\startdata
MW mass (I) & -- & -- \\
\hline
$+$ MW concentration (II) & 77.4 & 77.4 \\
\hline
\hline
$+$ LMC-mass subhalo (IIIa) & 57.1 & 44.2 \\
\hline
$+$ LMC recent accretion (IIIb) & 45.9 & 20.3 \\
\hline
$+$ LMC distance (IIIc) & 8.86 & 1.80 \\
\hline
\hline
$+$ GSE merger (IV) & 41.3 & 0.74 \\
\enddata
{\footnotesize \tablecomments{4455 MW-mass halos exist in \textsc{c125-1024}; $33$ pass all cuts.}\vspace{-3mm}}
\end{deluxetable}

Next, for the LMC, our $V_{\mathrm{max,sub}}$ cut is based on measurements of the LMC's rotation curve \citep{vanderMarel13054641}. Although we do not place an additional cut on the LMC's halo mass, we will show that our LMC analogs' halo masses are broadly consistent with LMC dynamical mass measurements (see Section~\ref{sec:lmc_chars}).\footnote{\cite{BoylanKolchin09114484} find that $\approx 10\%$ to $30\%$ of MW-mass halos host a subhalo with $M_{\mathrm{sub}}\gtrsim 10^{11}~\msun$. Some LMC analogs with $M_{\mathrm{sub}}<10^{11}~\msun$ satisfy our $V_{\mathrm{max,sub}}>55~\kms$ criterion, which explains why a larger fraction of our hosts pass this cut than in \cite{BoylanKolchin09114484}; see Table~\ref{tab:1}.} We impose a cut on $a_{\mathrm{infall}}$ so that our LMC analogs' infall times are consistent with those derived from proper-motion measurements \citep{Kallivayalil13010832}. Most of the LMC analogs are on first infall; see Section~\ref{sec:sample} for details on the final sample.
While the LMC's current $50~\kpc$ distance from the MW is measured to within $\pm 1~\kpc$ \citep{Pietrzynski2019}, we include halos with LMC analogs within $\pm 20~\kpc$ of this distance at $z=0$ in order to obtain a statistically representative sample of LMC analogs. The LMC distance criterion is our most restrictive cut and yields LMC analogs that are broadly consistent with the LMC's inferred orbit; we discuss this further in Section~\ref{sec:lmc_chars}.

Finally, our GSE selection criteria are less restrictive because observational uncertainties on its measured properties are large. For example, \cite{Helmi_streams} quotes an $\mathcal{O}(1~\Gyr)$ uncertainty on GSE's infall time, and an order-of-magnitude uncertainty on its stellar mass; additional modeling is required to infer GSE's halo mass (e.g., \citealt{Naidu210303251}). Given our broad ranges of allowed GSE parameters described in Section~\ref{sec:criteria}, about half of the hosts with recently accreted, nearby LMCs also have GSE analogs.

\subsection{Milky Way-est Sample}
\label{sec:sample}

We find that 33 halos satisfy the above criteria, out of 4455 halos in the MW mass range in \textsc{c125-1024}; the percentage of MW-mass halos in \textsc{c125-1024} that pass each cut is summarized in Table~\ref{tab:1}. We then identify hosts for which zoom-in simulations are computationally feasible based on the volume of their high-resolution regions: There are 25 such hosts, and we conduct zoom-in simulations for these. In five of these resimulations, the LMC analog's orbit noticeably differs from the parent box (potentially due to the butterfly effect; \citealt{Genel180707084}), so we do not analyze these hosts further. The resulting 20 zoom ins constitute the Milky Way-est suite.\footnote{One of these hosts (Halo327) was also in the Symphony MW sample \citep{MWW2015,Nadler_2023}; when comparing the two suites, we only include it in Milky Way-est.\label{foot:halo327}} Figure~\ref{fig:mwest_grid} shows the projected dark matter density within each host halo's virial radius at $z\approx 0$, arranged by host mass and concentration; each LMC analog halo is visible as a prominent substructure near its host's center.

Given the $\pm 20~\kpc$ range allowed for our LMC analogs' $z=0$ distances, we define the snapshot at which to analyze each Milky Way-est system, $a_{\mathrm{LMC,50}}$, as the time that its LMC analog is closest to $50~\kpc$ from the host center.\footnote{In one case (Halo659), we select the LMC's pericenter snapshot because pericenter occurs when the LMC reaches a distance of $44~\kpc$, which is sufficiently close to $50~\kpc$ for our analysis.} If there are multiple snapshots at which the LMC analog's distance is smaller than $50~\kpc$, we choose the earliest time. Because our LMC analogs are constrained to accrete recently, most LMC analogs reach their first pericenter at or near this point, with the exception of two LMC analogs on second infall (see Section~\ref{sec:lmc_chars} for details). In general, our LMC analogs' orbits do not exactly match those in the parent box; in order to ensure that all LMC analogs reach an appropriate distance, some Milky Way-est simulations are analyzed slightly after $z=0$. The resulting Milky Way-est analysis snapshots are typically within $\approx 500~\Myr$ of $z=0$, and are listed in Appendix~\ref{sec:host_properties}. We refer to these as the $z\approx 0$ snapshots, and we normalize scale factors for each Milky Way-est host relative to $a_{\mathrm{LMC,50}}$.

Although we do not explicitly constrain hosts' accretion histories at intermediate times, nearly half of the hosts in our sample accrete a subhalo with a peak mass above $10^{11}~\msun$ that merges between GSE disruption and LMC infall, potentially analogous to Sgr; we leave a detailed study of these systems to future work. The other half of our hosts are thus quiescent at intermediate times. We have also explored SMC analogs in Milky Way-est. However, no halos that satisfy all cuts above have a clear SMC analog, i.e., a massive subhalo on a long-lived orbit around the LMC. Finally, we note that Milky Way-est halos are not explicitly selected to reside in environments representative of the Local Group or Local Volume; as a result, massive neighboring host halos reminiscent of M31 surround a handful of Milky Way-est systems. Studying the impact of Milky Way-est hosts' environments on their accretion histories and subhalo populations is therefore an interesting area for future work.

\subsection{Symphony MW Sample}
\label{sec:symphonysample}

Here, we briefly describe the details of the Symphony MW suite, which we use in comparative analyses with the Milky Way-est suite presented in this work.

\begin{figure}[t!]
\includegraphics[trim={0 0cm 0 0},width=0.49\textwidth]{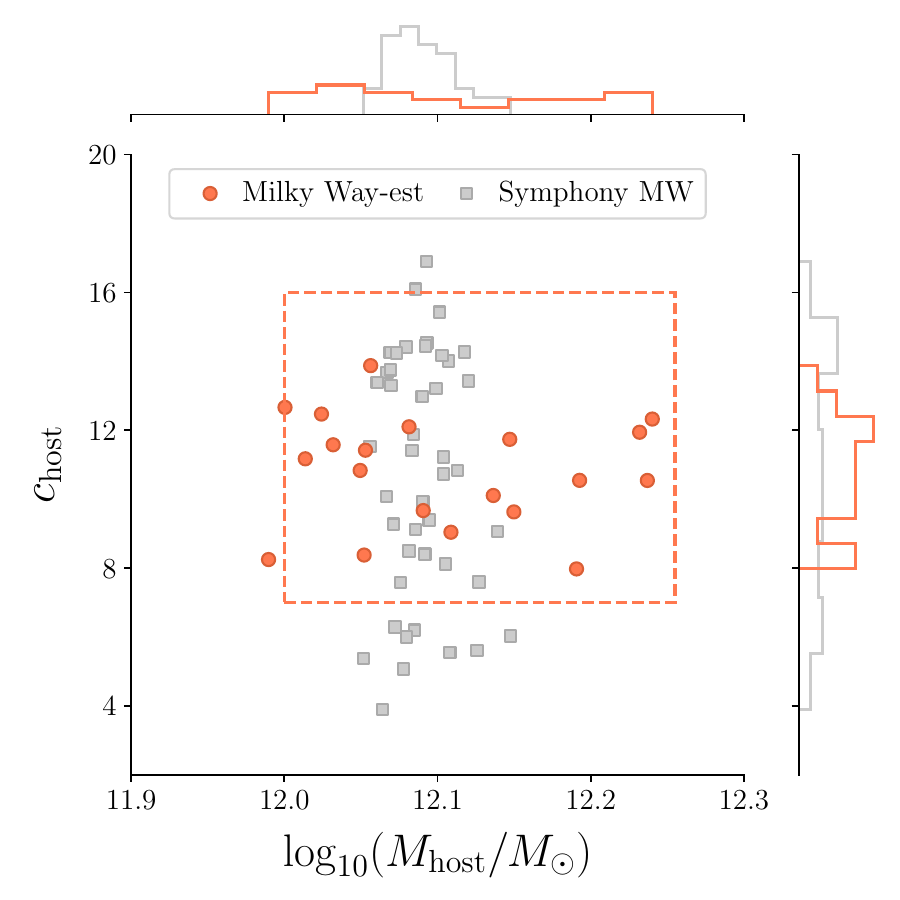}
\caption{Mass--concentration relation for Milky Way-est (orange circles) and Symphony MW hosts (gray squares). The box corresponds to our MW host halo selection cuts (i.e., criteria (I) and (II) in Section~\ref{sec:criteria}).}
\label{fig:host_mc_square}
\end{figure}

We use $44$ Symphony MW hosts from Nadler et al.\ (\citeyear{Nadler_2023}), i.e., the original sample of $45$ hosts minus Halo327; see footnote~\ref{foot:halo327}). These systems were first presented in \cite{MWW2015} and were selected from \textsc{c125-1024} to have a mass within $10^{12.09\pm0.02}~\msun$ at $z=0$ in the parent box. These hosts are subject to an isolation cut in the parent box, such that they cannot be within a distance of $4R_{\mathrm{vir,host}}$ from any more massive halo. Milky Way-est hosts are not subject to an explicit isolation cut. We observe in Figure~\ref{fig:host_mc_square} that Symphony MW hosts have a narrow mass distribution (due to the Symphony MW selection criteria) and wider concentration distribution than Milky Way-est hosts (due to our explicit concentration cut in Milky Way-est; criterion (II) from Table~\ref{tab:1}). To mitigate the differences in mass distributions, our subhalo population analyses are performed with subhalo masses normalized to the host mass to help isolate the effects of Milky Way-est selection criteria. Throughout this work, Symphony MW hosts are always analyzed at $z=0$.


\section{Properties of Milky Way-est Halos}
\label{sec:characteristics}

Having described the Milky Way-est sample, we now analyze the properties of the MW host halos (Section~\ref{sec:mw_chars}), LMC analog halos (Section~\ref{sec:lmc_chars}), and GSE analog halos (Section~\ref{sec:gse_chars}) in our zoom-in simulations.

\subsection{MW Analog Characteristics}
\label{sec:mw_chars}

After performing our zoom-in simulations, we identified host halos by matching their accretion histories to the MAHs of target hosts in \textsc{c125-1024}. Our hosts' MAHs match their counterparts in the parent box; for example, the average difference in $M_{\mathrm{host}}$ at $z\approx 0$ between our zoom ins and the parent box is $3\%$. In our zoom ins, all but one of the Milky Way-est hosts satisfy our MW halo concentration cut (Section~\ref{sec:criteria}), as shown by the orange circles in Figure~\ref{fig:host_mc_square}. Thus, these host halo properties are robust to the change in resolution from \textsc{c125-1024} to our zoom-in resimulations.

We also show Symphony MW halos (gray squares in Figure~\ref{fig:host_mc_square}), which span a narrower range of host mass and a broader range of concentration than the Milky Way-est host sample, given our observationally motivated cuts. When comparing Milky Way-est to Symphony MW in Section~\ref{sec:results}, we normalize subhalo masses by their corresponding host masses, which controls for the difference in the host mass distributions because subhalo abundances scale approximately linearly with host mass for MW-mass halos (e.g., \citealt{Nadler_2023}). At fixed host halo mass, subhalo abundances are also sensitive to host concentration (e.g., \citealt{MWW2015}) and other secondary properties (e.g., \citealt{Fielder180705180}), which in turn may be affected by the presence of LMC analogs in Milky Way-est hosts (e.g., \citealt{Fielder200702964}).

\subsection{LMC Analog Characteristics}
\label{sec:lmc_chars}

\begin{figure}[tb!]
\includegraphics[trim={0 0cm 0 0},width=0.49\textwidth]{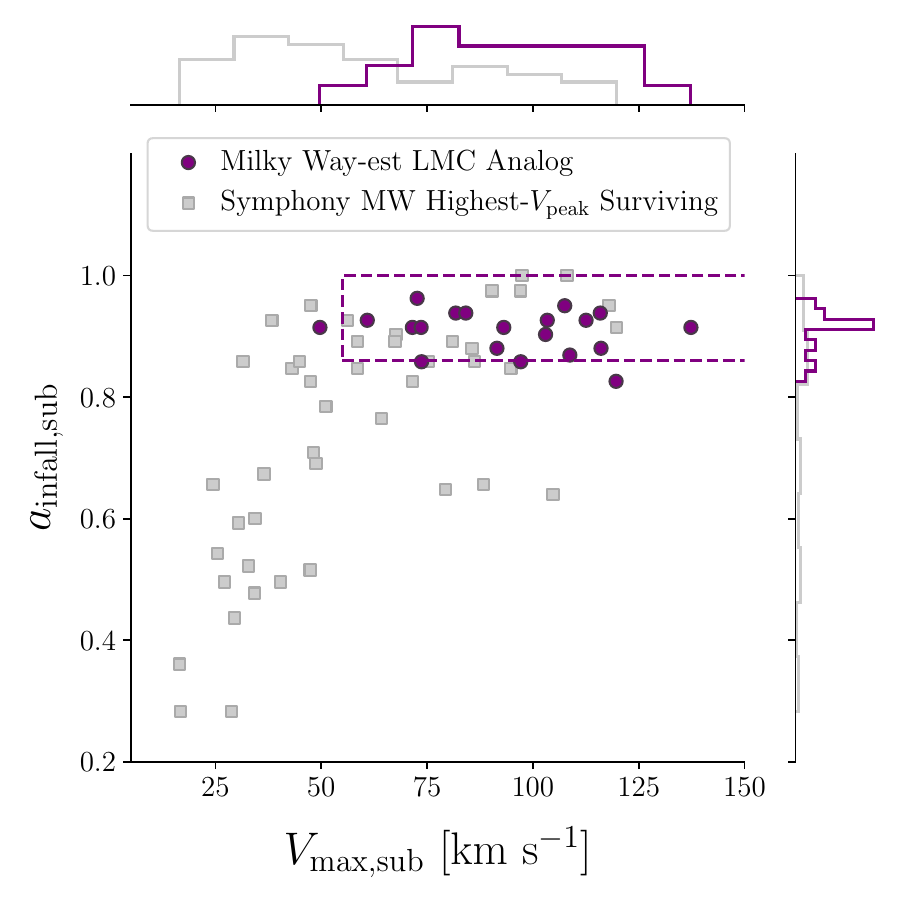}
\caption{Maximum circular velocity, $V_{\mathrm{max,sub}}$, vs. most recent infall time for Milky Way-est LMC analogs (purple circles) and the highest-$V_{\mathrm{peak,sub}}$ subhalo in each Symphony MW host (gray squares). The latest infall time is defined as the snapshot that a subhalo most recently accretes into its host. The box corresponds to our LMC analog selection cuts (i.e., criteria (IIIa) and (IIIb) in Section~\ref{sec:criteria}); Milky Way-est hosts are analyzed at the $z\approx 0$ snapshot when each LMC analog is nearest to a distance of $50~\kpc$ from the host center, consistent with selection criterion (IIIc).}
\label{fig:lmc_av_square}
\end{figure}

We identify LMC analogs by matching the orbits and MAHs of subhalos in our zoom-in simulations with \textsc{c125-1024}. All resulting LMC analogs match the halos in the parent box well; for example, their peak masses agree at the percent level. Figure~\ref{fig:lmc_av_square} shows the maximum circular velocity and accretion scale factor of each Milky Way-est LMC analog (purple circles).  
All Milky Way-est LMC analogs are most recently accreted within the last $2~\Gyr$ ($a_{\mathrm{infall}}\geq 0.86$, consistent with our selection criterion from Section~\ref{sec:criteria}), and almost all of these LMC analogs are on first infall. Furthermore, all of our LMC analogs have $V_{\mathrm{max,sub}}$ and $a_{\mathrm{infall}}$ values that satisfy our criteria from Section~\ref{sec:criteria} at the percent level. At the snapshots when our LMC analogs are closest to $50~\kpc$ from their respective hosts' centers, their distance distribution has a mean and host-to-host standard deviation of $r_{\mathrm{LMC}}=52\pm 11~\kpc$. Milky Way-est LMC analogs have a mean $V_{\mathrm{max,sub}}$ ($M_{\mathrm{sub}}$) of $94~\kms$ ($1.8\times 10^{11}~\msun$) with a standard deviation of $22~\kms$ ($1.4\times 10^{11}~\msun$). Of our 20 LMC analogs, 13 have present-day virial masses $M_{\mathrm{sub}}>10^{11}~\msun$ and are thus broadly consistent with measurements of the LMC’s current dynamical mass (e.g., \citealt{Erkal181208192,Shipp210713004}). We note that our LMC analog sample includes seven systems with $M_{\mathrm{sub}}<10^{11}~\msun$ because we only impose a $V_\mathrm{max,sub}$ cut (criterion IIIa from from Section~\ref{sec:criteria}; see Table~\ref{tab:sims} for a list of our LMC analogs' masses).

For comparison, Figure~\ref{fig:lmc_av_square} shows the same quantities for the subhalo with the highest $V_{\mathrm{peak,sub}}$ in each Symphony MW simulation (gray squares). We choose highest-$V_{\mathrm{peak,sub}}$ subhalos because they are expected to host the brightest surviving satellite of each Symphony MW host, as for our LMC analogs \citep{Lehmann151005651}. The highest-$V_{\mathrm{peak,sub}}$ Symphony MW subhalos are generally less massive than the Milky Way-est LMC analogs: In Symphony MW, the highest-$V_{\mathrm{peak,sub}}$ subhalos' mean $V_{\mathrm{max,sub}}$ ($M_{\mathrm{sub}}$) is $59~\kms$ ($6.8\times 10^{10}~\msun$). We find that there is a correlation between $V_{\mathrm{max,sub}}$ and $a_{\mathrm{infall}}$ for the highest-$V_{\mathrm{peak,sub}}$ Symphony MW subhalos, which is reasonable since earlier-accreted subhalos are less massive at infall and subsequently more stripped (e.g., \citealt{Kravtsov0401088}). In Milky Way-est, this correlation is not present because all LMC analogs are accreted recently.

\begin{figure}[tb!]
\hspace{-3mm}
\includegraphics[width=0.55\textwidth]{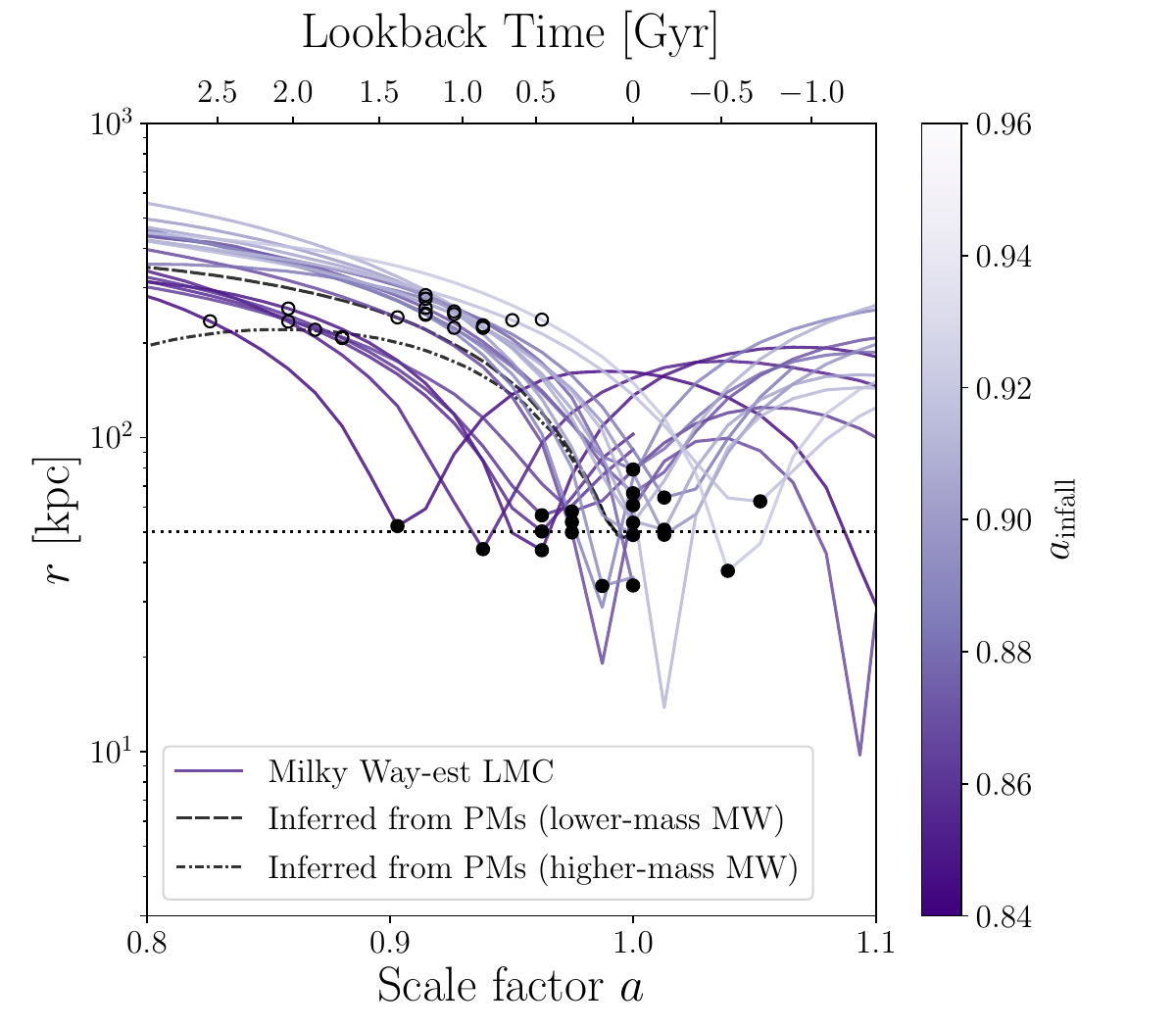}
\caption{Orbits of the LMC analogs in our Milky Way-est simulations (purple solid). Black filled points mark the $z\approx 0$ snapshot for each host used in our analyses, black unfilled points mark $a_{\mathrm{infall}}$, and line colors correspond to $a_{\mathrm{infall}}$. The horizontal dotted line marks $50~\kpc$. Additionally, solutions for the observed LMC's orbit based on proper-motion (PM) measurements are shown in black, from \cite{Patel200101746}; their fiducial LMC with a lower-mass (higher-mass) MW host is shown in dashed (dotted--dashed). Note that all LMC analogs are on first infall, except for Halo229 and Halo659, which have first pericenters at distances of $142~\kpc$ and $70~\kpc$, respectively, both at $a=0.59$.}
\label{fig:orbits}
\end{figure}

Figure~\ref{fig:orbits} shows the orbits of all Milky Way-est LMC analogs. As noted above, all but two of our LMC analogs are near their first pericenter at the $z\approx 0$ snapshot; after infall, their orbits are generally consistent with those inferred from proper-motion measurements \citep{Kallivayalil13010832}, including recent studies using Gaia data \citep{Vasiliev230409136}. 
We note that the orbits of our two LMC analogs on their second pericenter are also potentially consistent with orbits reconstructed from current LMC proper-motion measurements (e.g., see \citealt{Vasiliev230604837}).

\subsection{GSE Analog Characteristics}
\label{sec:gse_chars}

We identify GSE analogs by comparing the MAHs and orbits of disrupted subhalos in our zoom-in simulations to the GSE analog for each host in the parent box, following our procedure for LMC analogs in Section~\ref{sec:lmc_chars}. If multiple disrupted subhalos satisfy the GSE selection criteria in the parent box, we compare to the subhalo with the highest $M_{\mathrm{sub}}/M_{\mathrm{host}}$ when it reaches its peak mass. For most Milky Way-est hosts, we identify a GSE analog with an MAH and orbit that matches its counterpart in \textsc{c125-1024}. In a few cases, the GSE analog's peak mass is similar to that in the parent box, but its orbit and disruption time differ. In one case, we cannot identify a counterpart of the original GSE analog in our zoom-in resimulation, but we still find a disrupted subhalo consistent with our GSE selection criteria.

\begin{figure}[tb!]
\includegraphics[trim={0 0cm 0 0},width=0.49\textwidth]{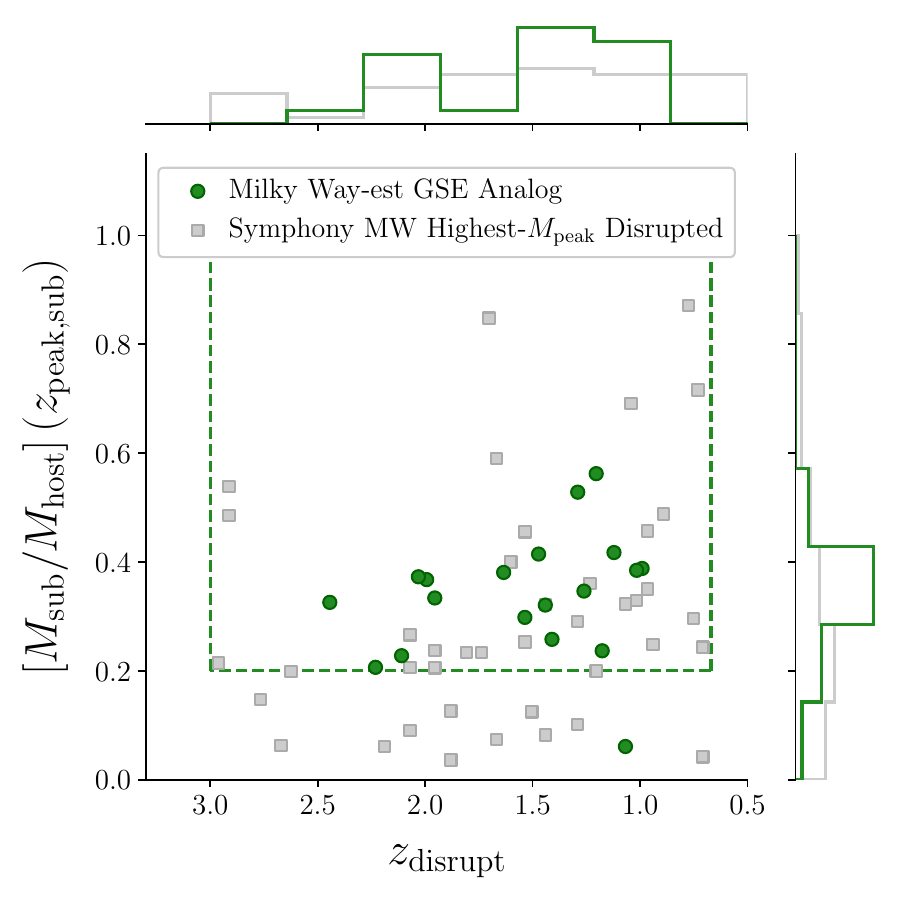}
\caption{Redshift at which a disrupted subhalo merges with the host, $z_{\mathrm{disrupt}}$, vs. the sub-to-host halo mass ratio for this object evaluated when its peak mass is achieved, $[M_{\mathrm{sub}}/M_{\mathrm{host}}](z_{\mathrm{peak,sub}})$. Points show Milky Way-est GSE analogs (green circles) and highest-$M_{\mathrm{peak,sub}}$ Symphony MW disrupted subhalos with $0.67<z_{\mathrm{disrupt}}<3$ (gray squares). The box corresponds to our GSE analog selection cuts (i.e., criterion (IV) in Section~\ref{sec:criteria}).}
\label{fig:adisrupt_massratio}
\end{figure}

Figure~\ref{fig:adisrupt_massratio} shows the disruption times and sub-to-host halo mass ratios for our Milky Way-est GSE analogs (green circles). We compare these to the most massive disrupted subhalo over the relevant range of $z_{\mathrm{disrupt}}$ in each Symphony MW simulation (gray squares). We evaluate sub-to-host halo mass ratios at the snapshot when the GSE analog (or highest-$M_{\mathrm{peak,sub}}$ disrupted subhalo) achieves its peak mass to match our GSE selection criteria. For all Milky Way-est hosts except the case noted above, the GSE analog is consistent with our constraints, shown by the dashed region in Figure~\ref{fig:adisrupt_massratio}.

Our GSE analogs merge with a mean $z_{\mathrm{disrupt}}$ of $1.54\pm 0.43$ (or mean $a_{\mathrm{disrupt}}$ of $0.40\pm 0.06$), where uncertainties represent the host-to-host standard deviation. This distribution is consistent with the Symphony MW highest-$M_{\mathrm{peak,sub}}$ disrupted subhalos we compare to, which were selected over the same disruption time range. The mean peak mass of Milky Way-est hosts' GSE analogs is $1.1\times 10^{11}~\msun$, versus $1.5\times 10^{11}~\msun$ for Symphony MW's highest-$M_{\mathrm{peak,sub}}$ disrupted subhalos. Despite these similar mass distributions, many Symphony MW disrupted subhalos have sub-to-host halo mass ratios \emph{below} our GSE criterion. This result is influenced by our LMC constraint---in particular, the LMC contributes significantly to Milky Way-est hosts' total halo masses at late times, such that Milky Way-est hosts have lower masses than average during the epoch of the GSE merger (see Section~\ref{sec:mah}). Thus, we measure larger sub-to-host halo merger mass ratios for Milky Way-est GSE analogs compared to Symphony MW highest-$M_{\mathrm{peak,sub}}$ disrupted subhalos, on average.

Despite this general trend, Symphony MW also contains disrupted subhalos with sub-to-host halo mass ratios $\gtrsim 0.5$. Such values are possible because this mass ratio is evaluated when the disrupted subhalo achieves $M_{\mathrm{peak,sub}}$. Before the merger, these objects may be stripped while the host grows, resulting in a sub-to-host halo mass ratio below $0.5$ when the merger occurs.


\section{Results}
\label{sec:results}

We now present our results, focusing on host halo MAHs (Section~\ref{sec:mah}), SHMFs (Section~\ref{sec:shmf}), subhalo radial distributions (Section~\ref{sec:shrd}), and subhalo spatial anisotropy (Section~\ref{sec:aniso}).

\subsection{Host Halo Formation Histories}
\label{sec:mah}

\begin{figure*}[t!]
\centering
\includegraphics[trim={0 0cm 0 0},width=\textwidth]{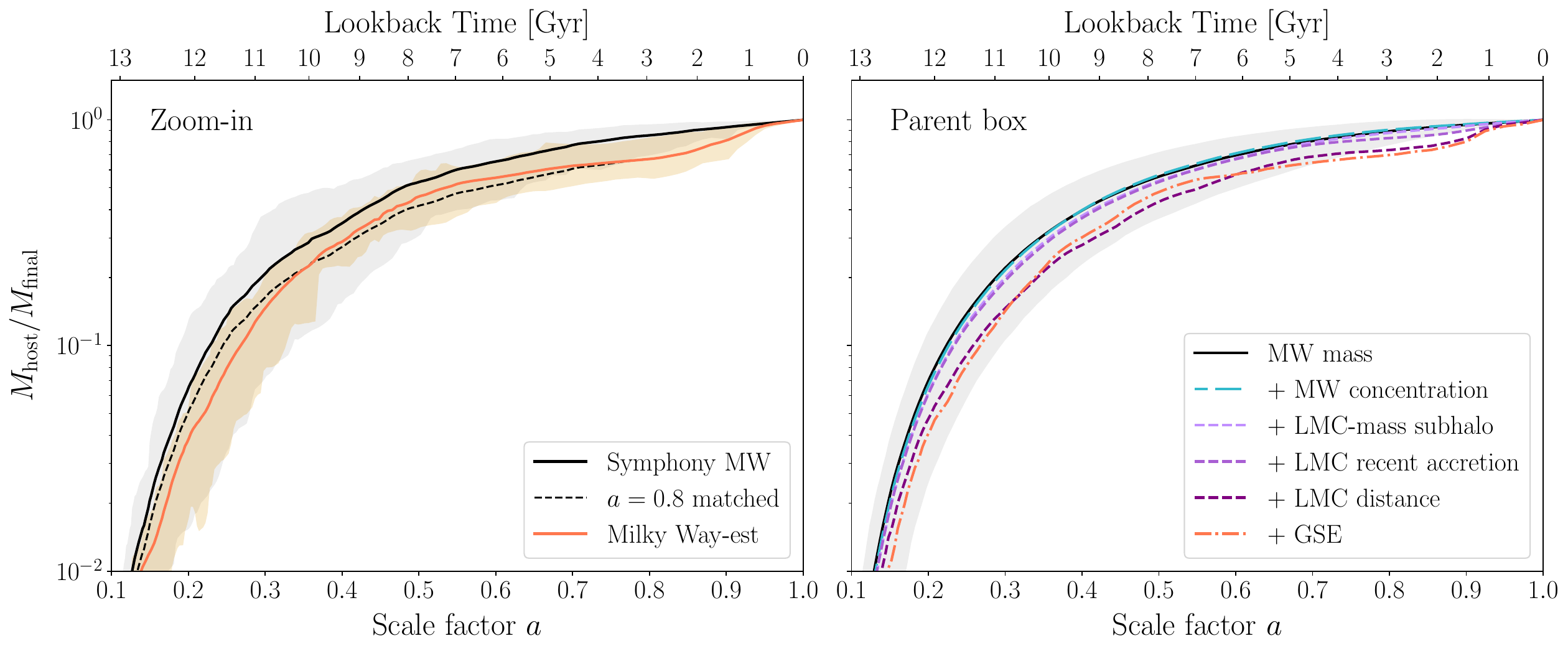}
\caption{Left panel: mean MAHs of Milky Way-est (gold) and Symphony MW (black solid) zoom-in hosts. The mean Symphony MW MAH normalized to the Milky Way-est value at $a=0.8$ is also shown (black dashed). Shaded regions show the 16th--84th MAH percentiles for each suite. Each host's mass is normalized to its $z=0$ value before stacking. Scale factors are normalized to the scale factor of each host's final snapshot before averaging. 
Right panel: mean MAHs of halos in \textsc{c125-1024} that satisfy successive Milky Way-est selection criteria (see Section~\ref{sec:criteria} and Table~\ref{tab:1}).  
The gray band corresponds to MW-mass halos, based on our Milky Way-est mass selection criterion, and colored lines show samples that satisfy successive cuts on host concentration (light blue long-dashed), LMC analog mass, infall time, and present-day distance (lightest to darkest purple short-dashed), and GSE analogs (gold dotted--dashed).}
\label{fig:mahs}
\end{figure*}

Our MW analog selection criteria systematically affect the formation histories of Milky Way-est host halos. To demonstrate this, the left panel of Figure~\ref{fig:mahs} shows host halo MAHs for Milky Way-est (gold) and Symphony MW (black), normalized to the mass of each host at $z\approx 0$ (or $z=0$ for Symphony MW). Milky Way-est hosts form later, on average, than Symphony MW hosts; for example, the mean scale factor at which half the final mass is achieved, $a_{1/2}$, is $0.57\pm 0.1$ for Milky Way-est and $0.49 \pm 0.1$ for Symphony MW, where uncertainties represent the host-to-host standard deviation. This $\approx 1\sigma$ difference relative to the host-to-host scatter is significant given the $\pm 0.01$ statistical error on our $a_{1/2}$ measurements, estimated using jackknife resampling, and corresponds to a $1.3~\Gyr$ shift in average half-mass formation time. In turn, the slope of the mean Milky Way-est MAH differs from Symphony MW at late times, when our LMC analog subhalos accrete.

Before interpreting these results, we describe a few technical aspects of our MAH calculation. First, we present MAHs using host virial masses reported by \textsc{Rockstar}. This includes the masses of all subhalos within the virial radius at a given time, although we note that the late-time growth described above for Milky Way-est hosts persists even if subhalos' masses are subtracted from the host mass at each snapshot. Second, a small subset of Milky Way-est systems have LMC analogs that are particularly massive (e.g.\ $M_{\mathrm{sub}}> 1.5\times 10^{11} M_{\odot}$ at $z\approx 0$). However, the shift toward later assembly for Milky Way-est hosts persists even when these systems are excluded. Finally, we have checked that additional differences in host selection criteria do not drive this MAH shift. In particular, host samples from \textsc{c125-1024} selected with and without the ``secondary'' Symphony MW host selection cuts on environment and Lagrangian volume \citep{Nadler_2023} display consistent MAHs, implying that these characteristics do not drive the MAH difference.

To unpack these findings, we study the impact of each Milky Way-est selection criterion on the MAHs of host halos in \textsc{c125-1024}. As shown in the right panel of Figure~\ref{fig:mahs}, the cut on host concentration (criterion II in Section~\ref{sec:criteria}) does not significantly impact hosts' normalized MAHs. The existence of a massive subhalo (criterion IIIa from Section~\ref{sec:criteria}) also does not have a significant impact; in general, such subhalos do not accrete as recently as our LMC analogs and thus may be heavily stripped by $z=0$. Instead, the shift toward more recent assembly is largely driven by the existence of a recently accreted and nearby LMC analog halo (criteria IIIb and IIIc from Section~\ref{sec:criteria}), consistent with the findings in \cite{Lu160502075}. We note that LMC analogs in the parent box that satisfy our $V_{\mathrm{max,sub}}$, $a_{\mathrm{infall}}$, and distance cuts are $\approx 50\%$ more massive than subhalos that satisfy our $V_{\mathrm{max,sub}}$ and $a_{\mathrm{infall}}$ criteria but are not on LMC-like orbits. This is reasonable, as massive subhalos sink to the host center more quickly due to dynamical friction \citep[e.g.,][]{vandenBosch151001586}.

Adding our GSE analog cuts (criterion IV from Section~\ref{sec:criteria}) produces a slight ``bump'' in the normalized MAHs during the epoch of the GSE merger; however, this feature is only marginally significant and lies within the host-to-host scatter. In particular, at $a=0.5$, the ``$+$~GSE'' (``$+$~LMC distance'') sample in Figure~\ref{fig:mahs} has a mean normalized mass of $0.51$ ($0.46$). For comparison, the jackknife uncertainty on the mean MAH is $0.03$, and the host-to-host standard deviation is $0.16$. The GSE bump is followed by a flatter normalized MAH because the GSE analog represents one of the largest early accretion events for each Milky Way-est host. This flattening makes the late-time increase in the normalized MAH due to the LMC analog even more dramatic in Milky Way-est, compared to typical Symphony MW hosts.

\begin{figure*}[t!]
\includegraphics[trim={0 0cm 0 0},width=0.49\textwidth]{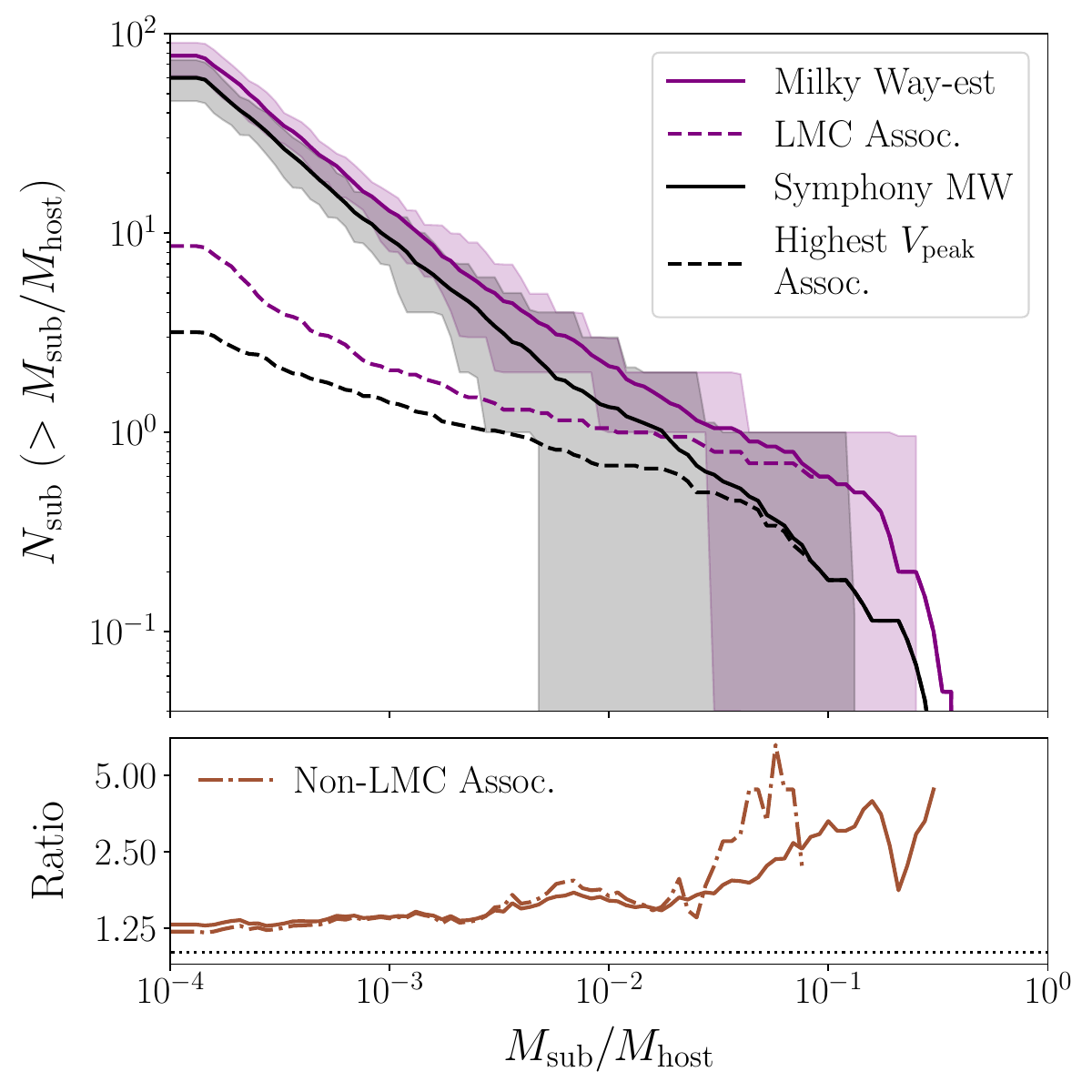}
\includegraphics[trim={0 0cm 0 0},width=0.49\textwidth]{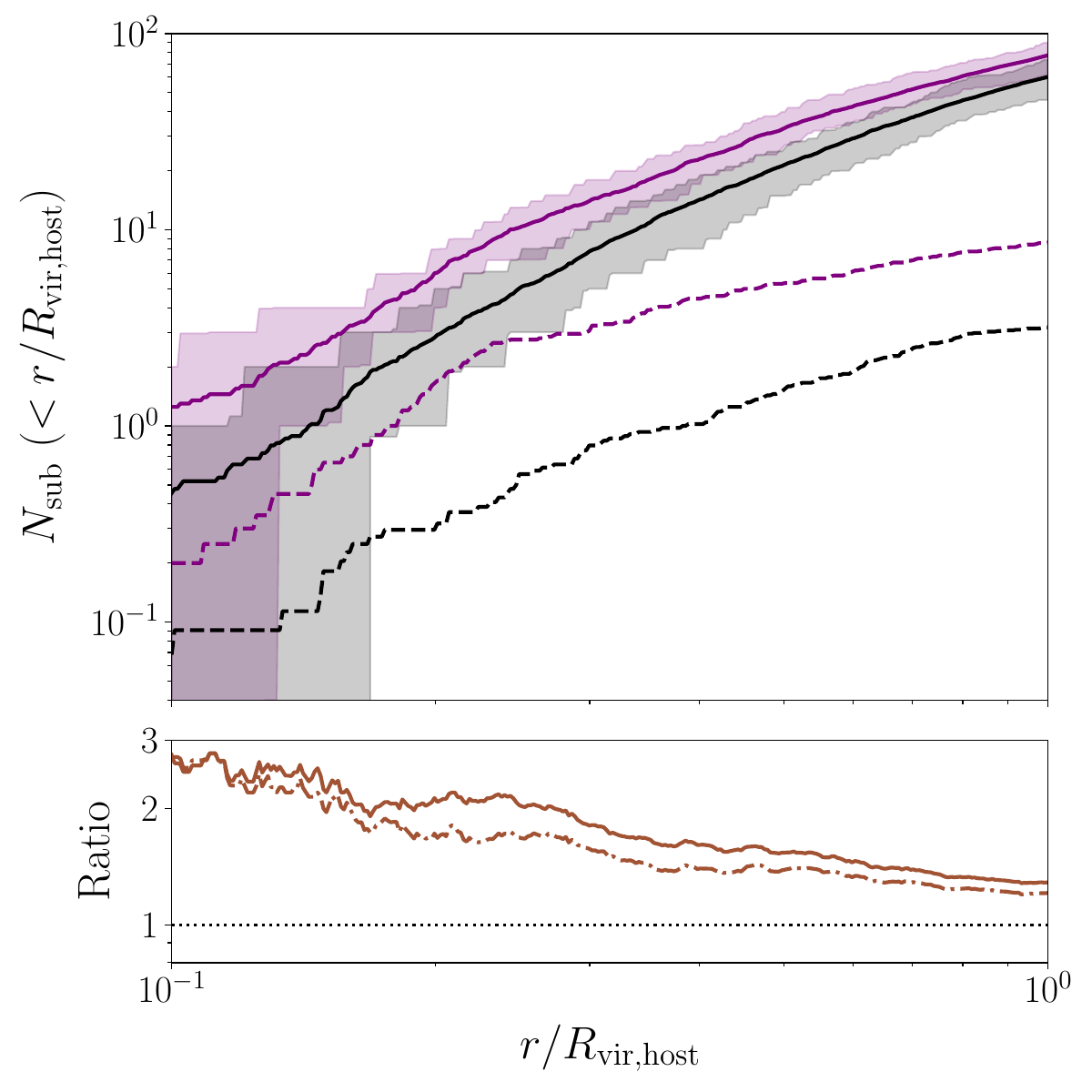}
\caption{Left panel:
mean SHMFs, measured using present-day subhalo virial mass ($M_{\mathrm{sub}}$) and normalized by host virial mass ($M_{\mathrm{host}}$), for all subhalos within the virial radius of Milky Way-est (solid purple) and Symphony MW (solid black) hosts. Dashed lines show the contribution to these mass functions from subhalos associated with the LMC analogs in Milky Way-est and the highest-$V_{\mathrm{peak,sub}}$ subhalos in Symphony MW. Right panel: mean subhalo radial distributions in Milky Way-est and Symphony MW systems as a function of distance from the host halo center, in units of $R_{\mathrm{vir,host}}$. 
Dashed lines show the radial distributions of subhalos associated with the LMC analog or highest-$V_{\mathrm{peak,sub}}$ subhalo, as in the left panel. In both panels, shaded bands show 16th--84th percentiles, and lower panels show the ratio of Milky Way-est to Symphony MW measurements for all subhalos (solid) and non-LMC-associated subhalos (dotted--dashed), with black dotted lines marking a ratio of 1.}
\label{fig:shmf_raddistri}
\end{figure*}

We also compare Milky Way-est and Symphony MW host MAHs up to (i) $a=0.25$ ($z=3$; before our GSE analogs accrete) and (ii) $a=0.8$ ($z=0.25$; before our LMC analogs accrete; see the black dashed line in Figure~\ref{fig:mahs}) to examine how these constraints affect Milky Way-est hosts' MAHs at other times. We find that hosts' specific accretion rates are similar between the suites for $a<0.25$. At $a=0.25$, Milky Way-est hosts are $39\%$ less massive, on average, with a mean mass of $1.0\times 10^{11}~\msun$ versus $1.6\times 10^{11}~\msun$ for Symphony MW. During GSE infall and disruption ($0.25<a_\mathrm{disrupt}<0.6$ or $0.67<z_\mathrm{disrupt}<3$), Milky Way-est hosts' specific accretion rates are higher than in Symphony MW, although Milky Way-est hosts are still less massive at these times, on average. 
At later times ($a>0.8$), and particularly during LMC infall, Milky Way-est hosts overtake Symphony MW hosts in terms of both specific and absolute growth.

\subsection{Subhalo Mass Functions}
\label{sec:shmf}

The left panel of Figure~\ref{fig:shmf_raddistri} shows mean SHMFs for Milky Way-est (purple) and Symphony MW (black). SHMFs are measured using present-day virial mass and normalized to each host's virial mass at $z\approx 0$ (or at $z=0$ for Symphony MW). Milky Way-est systems host a mean number of $77\pm 13$ subhalos with $M_{\mathrm{sub}}/ M_{\mathrm{host}}>1.4\times 10^{-4}$ (or $M_{\mathrm{sub}}\gtrsim 10^{8}~\msun$), versus $60\pm 13$ for Symphony MW, where uncertainties represent the host-to-host standard deviation. Thus, at low masses the amplitude of the mean Milky Way-est SHMF is $22\%$ higher than in Symphony MW, while the SHMF slopes are consistent. This difference is statistically significant given the $\approx 3\%$ Poisson error on our stacked SHMF measurement and represents a $\approx 1.5\sigma$ difference relative to the host-to-host scatter. At higher masses ($M_{\text{sub}} / M_{\text{host}}\gtrsim 10^{-2}$), LMC analogs enhance Milky Way-est SHMFs relative to Symphony MW. This follows because the highest-$V_{\mathrm{peak,sub}}$ subhalos in Symphony MW are roughly 1 order of magnitude less massive than our LMC analogs, on average (see Section~\ref{sec:lmc_chars} and Figure~\ref{fig:lmc_av_square}).

\subsubsection{LMC-associated Subhalos}

The SHMF enhancement in Milky Way-est is partially due to subhalos accreted with our LMC analogs. To demonstrate this, the dashed lines in the top-left panel of Figure~\ref{fig:shmf_raddistri} show the contribution to the total SHMFs from the LMC analogs in Milky Way-est, and from the highest-$V_{\mathrm{peak,sub}}$ subhalos in Symphony MW. We identify LMC-associated subhalos within the LMC analog's virial radius at the time step before LMC accretion into the MW virial radius, following \cite{Nadler191203303}. Comparing the mean SHMFs, we find that Milky Way-est LMC analogs contribute a mean of $9\pm 6$ subhalos down to our $M_{\mathrm{sub}}/M_{\mathrm{host}}>1.4\times10^{-4}$ resolution limit, where uncertainties again represent the host-to-host standard deviation. Meanwhile, subhalos associated with the highest-$V_{\mathrm{peak,sub}}$ systems in Symphony MW only contribute $3\pm 3$ subhalos down to the same limit. The dotted--dashed line in the bottom-left panel of Figure~\ref{fig:shmf_raddistri} shows the SHMF ratio excluding LMC (or highest-$V_{\mathrm{peak,sub}}$)-associated subhalos and confirms that the difference is not entirely due to subhalos within the LMC's virial radius when it accretes.

Thus, $\approx 50\%$ of the difference between Milky Way-est and Symphony MW SHMF amplitudes can be directly attributed to LMC-associated subhalos. We also repeat our calculation using LMC-associated subhalos within $2\rvir$ of the LMC before that region overlaps with each Milky Way-est host's virial radius. This increases the mean abundance of LMC-associated subhalos in Milky Way-est from 9 to 18 and explains the remaining difference in SHMF amplitudes, indicating that the remaining excess subhalos in Milky Way-est also fell in from a spatial region associated with the LMC.

\begin{figure}[t!]
\hspace{-4mm}
\includegraphics[width=0.49\textwidth]{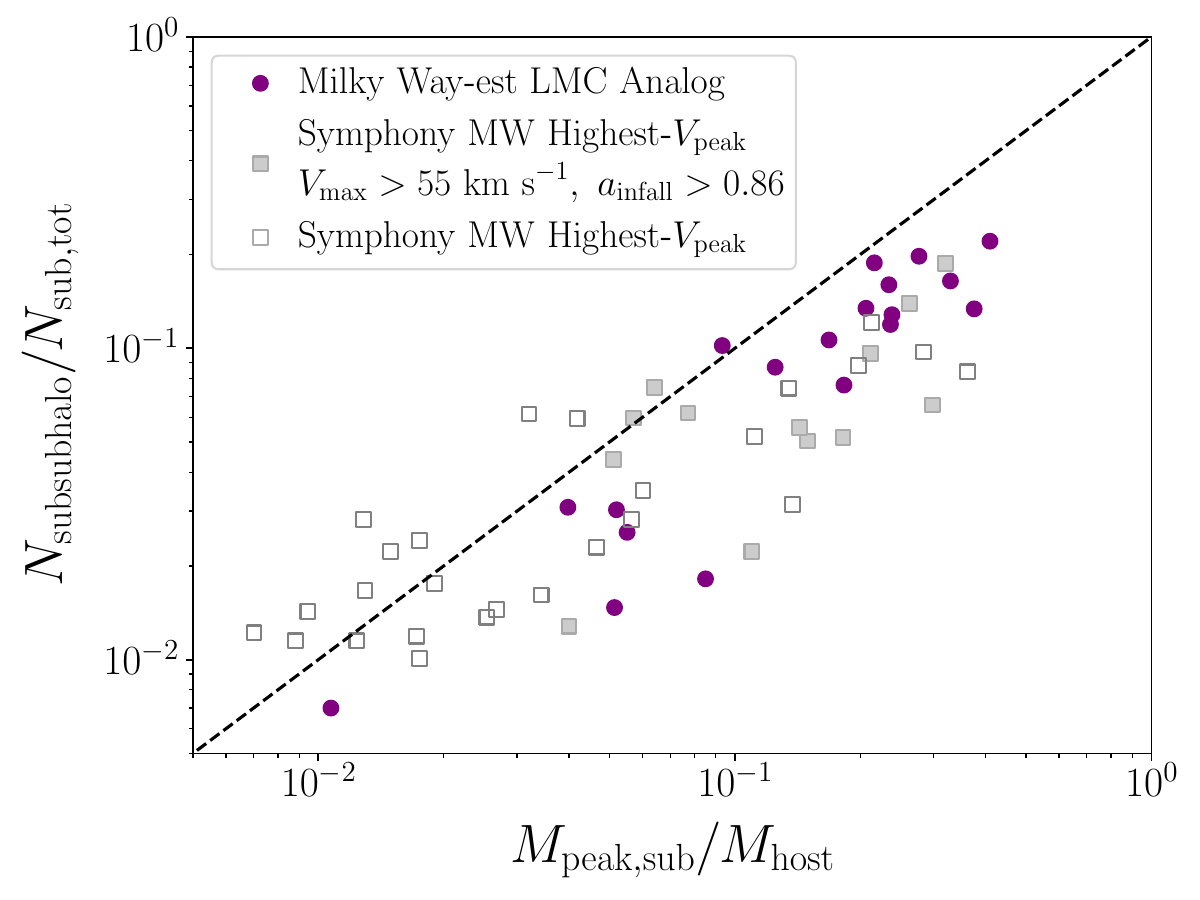}
\caption{Fraction of subhalos with $M_{\mathrm{sub}}>1.2\times 10^8~\msun$ contributed by Milky Way-est LMC analogs (purple circles) and Symphony MW highest-$V_{\mathrm{peak,sub}}$ subhalos (gray squares) vs. $M_{\mathrm{peak,sub}}/M_{\mathrm{host}}$ for the LMC analog or highest-$V_{\mathrm{peak,sub}}$ subhalo. Filled (unfilled) gray squares show Symphony MW highest-$V_{\mathrm{peak,sub}}$ subhalos that pass (do not pass) our LMC analog $V_{\mathrm{max,sub}}$ and $a_{\mathrm{infall}}$ cuts. The dashed black line shows the one-to-one relation.}
\label{fig:frac_associated}
\end{figure}

Figure~\ref{fig:frac_associated} shows the fraction of each host's total subhalo population with $M_{\mathrm{sub}}>1.2\times10^{8}~\msun$ contributed by Milky Way-est LMC analogs (purple circles) and Symphony MW highest-$V_{\mathrm{peak,sub}}$ subhalos (unfilled and filled gray squares). This fraction scales roughly linearly with $M_{\mathrm{peak,sub}}/M_{\mathrm{host}}$, which is reasonable given that subhalo abundance scales linearly with host mass (e.g., \citealt{Nadler_2023}). Subhalos associated with Milky Way-est LMC analogs constitute a mean of $(10\pm 6)\%$ of the total subhalo population down to $M_{\mathrm{sub}}>1.2\times10^{8}~\msun$, versus $(4\pm 4)\%$ for systems associated with Symphony MW highest-$V_{\mathrm{peak,sub}}$ subhalos, where uncertainties represent the host-to-host standard deviation. This $\approx 1\sigma$ difference follows because, on average, our LMC analogs are more massive and accrete more recently than Symphony MW highest-$V_{\mathrm{peak,sub}}$ subhalos. 
In particular, Symphony MW highest-$V_{\mathrm{peak,sub}}$ subhalos within our LMC analogs' range of $a_{\mathrm{infall}}$ and $V_{\mathrm{max,sub}}$ contribute $(7\pm 5)\%$ of the total subhalo population, on average (filled gray squares in Figure~\ref{fig:frac_associated}), which is still slightly lower than the LMC-associated subhalo contribution in Milky Way-est. 
It will therefore be interesting to study how SHMF enhancement depends on our LMC analogs' orbits \citep{Patel200101746,DSouza210413249,Barry230305527}.

\subsubsection{GSE-associated Subhalos}

Subhalos accreted with our GSE analogs may also contribute to the difference in SHMFs (see, e.g., \citealt{Bose190904039}). By identifying the number of subhalos within each GSE analog's virial radius at the time of accretion, we find a mean of $3\pm 3$ surviving GSE-associated subhalos with $M_{\mathrm{sub}}>1.2\times 10^{8}$ in Milky Way-est hosts at $z\approx 0$, on average, where uncertainties represent the host-to-host standard deviation. This contribution is small compared to the LMC-associated subhalos discussed above. However, we caution that it is difficult to robustly track subhalos associated with our GSE analogs throughout their mergers and over the subsequent $\approx 10~\mathrm{Gyr}$ using {\sc Rockstar} and {\sc Consistent-Trees}. We leave a more detailed analysis of GSE-associated subhalos using particle tracking, which can follow subhalos much longer (e.g., \citealt{Mansfield230810926}), to future work.

\begin{figure*}[t!]
\hspace{-0.5cm}
\includegraphics[trim={0 1cm 0 0},width=\textwidth]{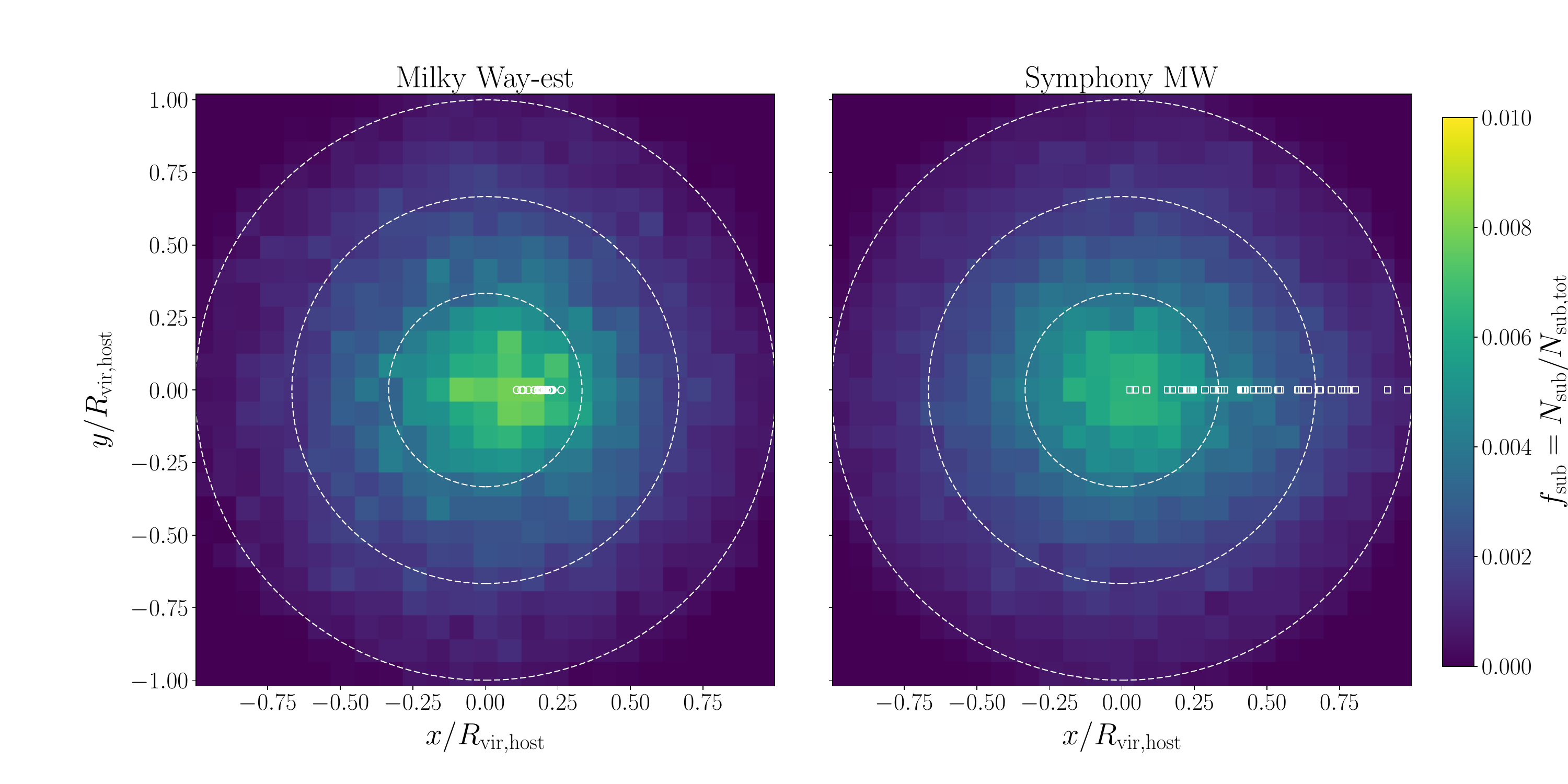}
\caption{Projected maps of subhalo counts stacked over all Milky Way-est (left panel) and Symphony MW (right panel) hosts. Distances are measured in units of each host's virial radius before stacking. Subhalo counts are normalized by the total number of subhalos in each host, such that each pixel shows the fraction of subhalos in a given projected area. Subhalo populations are rotated in three dimensions such that the center of each Milky Way-est LMC analog (white circles) or Symphony MW highest-$V_{\mathrm{peak,sub}}$ subhalo (white squares) lie along the $x$-axis. White dashed rings correspond to $1/3~R_{\mathrm{vir,host}}$, $2/3~R_{\mathrm{vir,host}}$, and $R_{\mathrm{vir,host}}$.}
\label{fig:anisotropy}
\end{figure*}

To indirectly assess the impact of GSE-associated subhalos on the SHMF, we split the Milky Way-est sample by the median peak mass ($\approx 10^{11}~\msun$) of the largest subhalo that merges with the host in the redshift interval $0.67<z<3$, corresponding to our GSE disruption window (Section~\ref{sec:criteria}). SHMFs for these subsamples are similar, consistent with our finding that GSE-associated subhalos contribute negligibly to Milky Way-est SHMFs at $z=0$, given our analysis tools and resolution cut.

\subsubsection{Summary}

We have shown that Milky Way-est systems host $22\%$ more subhalos with $M_{\mathrm{sub}}\gtrsim 10^8~\msun$, on average, than Symphony MW hosts that are not selected to satisfy our GSE and LMC analog criteria. GSE-associated subhalos contribute negligibly to this difference. Subhalos within the virial radius of the LMC at the time of LMC infall explain half of the difference, and subhalos within $2\rvir$ of the LMC when that region is entirely outside of the host explain the remaining difference. We have confirmed that Milky Way-est and Symphony MW SHMFs are similar at $a=0.8$ ($z=0.25$), consistent with these results. Finally, we note that secondary host properties are unlikely to affect this interpretation; in particular, although host concentration correlates with subhalo abundance in general \citep{MWW2015}, the difference in SHMF amplitude persists when excluding Symphony MW hosts outside of the Milky Way-est host concentration range.

\subsection{Subhalo Radial Distributions}
\label{sec:shrd}

Subhalo radial distributions are shown in the top-right panel of Figure~\ref{fig:shmf_raddistri}; we measure distances in units of each host's virial radius and apply our $M_{\mathrm{sub}}/M_{\mathrm{host}}>1.4\times10^{-4}$ cut. On average, Milky Way-est radial distributions are enhanced relative to Symphony MW at $r/R_{\mathrm{vir,host}}\lesssim 0.5$ ($r\lesssim 150~\mathrm{kpc}$). This is particularly noticeable in the innermost regions, $r/R_{\mathrm{vir,host}}\lesssim 0.3$ ($r\lesssim 100~\mathrm{kpc}$), where mean subhalo abundances are enhanced by $81\%$ in Milky Way-est relative to Symphony MW.

The enhancement of Milky Way-est inner radial distributions is partially driven by our LMC analogs. At $z\approx 0$, each LMC analog is roughly $50~\mathrm{kpc}$ from the host center; for a typical Milky Way-est host with $R_{\mathrm{vir,host}}=300~\kpc$, this translates to $r/R_{\mathrm{vir,host}}\approx 0.2$. Within this distance, LMC-associated subhalos constitute $27\%$ of the mean Milky Way-est subhalo abundance, compared to $10\%$ within the entire virial radius, on average. At larger distances, this contribution of LMC-associated subhalos decreases. This can be understood because (before the LMC system's first pericenter) LMC-associated subhalos are mostly contained within the LMC analog's virial radius of $\approx 150~\kpc$, which extends to $\approx 200~\kpc$ from the host center.

Thus, subhalos directly associated with the LMC contribute to the enhancement of the inner radial distribution in Milky Way-est relative to Symphony MW but do not entirely explain the difference between the suites. The dotted--dashed line in the bottom-right panel of Figure~\ref{fig:shmf_raddistri} shows the radial distribution ratio excluding LMC (or highest-$V_{\mathrm{peak,sub}}$)-associated subhalos, and (as for the SHMF) again confirms that the difference is not entirely due to the LMC. We note that subhalos within $2\rvir$ of each LMC analog when that region is outside of the host account for most of the remaining difference within $100~\kpc$. Additionally, although GSE-associated subhalos are expected to preferentially be found in the inner regions of the host (e.g., \citealt{Bose190904039}), our tests in Section~\ref{sec:shmf} indicate that these systems do not significantly contribute to present-day Milky Way-est subhalo abundances given our analysis tools. It is therefore unlikely that the remaining radial distribution difference is explained by GSE-associated subhalos. As for the SHMF, future work that studies the origins of subhalos which contribute to the inner radial distribution difference will be useful.

\subsection{Spatial Anisotropy of Subhalo Populations}
\label{sec:aniso}

Milky Way-est LMC analogs accrete recently, accompanied by their own subhalos (see Section~\ref{sec:shmf}).

Because we analyze most LMC analogs at or near their first pericenter, we expect that LMC-associated subhalos are still clustered near their LMC analog host at $z\approx 0$ (e.g., \citealt{DSouza210413249}). This effect would induce spatial anisotropy in the subhalo population in excess of that for an ``average’' MW-mass halo, consistent with LMC-induced clustering in angular momentum space reported by previous studies (e.g., \citealt{Garavito-Camargo210807321,Garavito-Camargo231111359G}). To assess this, we measure the spatial anisotropy of subhalo populations by calculating the fraction of subhalos in the hemisphere that points from the host center to the LMC analog at the $z\approx 0$ snapshot in each Milky Way-est simulation; an anisotropy measure of $50\%$, therefore, corresponds to an isotropic subhalo distribution. In Symphony MW, we use the highest-$V_{\mathrm{peak,sub}}$ subhalo at $z=0$ to define the preferred direction.

We find that subhalos are preferentially located in the present-day direction of our LMC analogs and that this effect is strongest in the inner regions of Milky Way-est hosts. In particular, the mean anisotropy of Milky Way-est subhalo populations is $58\%$ within $1/3~R_{\mathrm{vir,host}}$, $56\%$ within $2/3~R_{\mathrm{vir,host}}$, and $54\%$ within $R_{\mathrm{vir,host}}$.
In Symphony MW, we observe little to no anisotropy, with $51\%$--$52\%$ of all subhalos found in the highest-$V_{\mathrm{peak,sub}}$ subhalo's hemisphere at all radii. For both suites, the host-to-host standard deviation of these measurements is $\approx 5\%$. These differences between Milky Way-est and Symphony MW are statistically significant, given the $\approx 3\%$ Poisson error on our stacked anisotropy measurements, and represent a $\approx 1\sigma$ difference relative to the host-to-host scatter.

To illustrate these results, Figure~\ref{fig:anisotropy} shows projected maps of subhalo counts in Milky Way-est and Symphony MW, normalized to the total number of subhalos in each host and stacked over all hosts within each suite. To produce these maps, we perform a three-dimensional rotation for each subhalo population such that the LMC analog (or highest-$V_{\mathrm{peak,sub}}$ subhalo) lies on the $x$-axis. Subhalos are clearly more centrally concentrated in Milky Way-est compared to Symphony MW, consistent with our findings in Section~\ref{sec:shrd}. Furthermore, in the inner regions of Milky Way-est hosts, bright squares (i.e., projected areas with higher fractional subhalo abundance) are found in the direction of the LMC analogs, consistent with our anisotropy measurements above.


\section{Discussion}
\label{sec:future}

We now discuss the implications of Milky Way-est subhalo populations (Section~\ref{sec:subhalo_implications}), future work that simultaneously models the impact of GSE and LMC on various observables (Section~\ref{sec:unusual}), and relevant caveats (Section~\ref{sec:future_work}).

\subsection{Milky Way-est Subhalo Populations}
\label{sec:subhalo_implications}

The $\approx 20\%$ enhancement in Milky Way-est SHMFs (Section~\ref{sec:shmf}) and the accompanying $\approx 80\%$ enhancement of the inner radial distribution (Section~\ref{sec:shrd}) and spatial anisotropy (Section~\ref{sec:aniso}) have important implications for theoretical modeling and analyses of near-field data.

\subsubsection{Comparison to Previous Work}
\label{sec:compare_to_prev}

In Milky Way-est systems, the mean number of LMC-associated subhalos above our resolution cut is $9\pm 6$, where uncertainties represent the host-to-host standard deviation. This prediction corresponds to $(10\pm 6)\%$ of the total subhalo population with $M_{\mathrm{sub}}>1.2\times 10^8~\msun$ and is fairly consistent with the number of LMC-associated satellite galaxies inferred from current observations of the MW satellite population (e.g., \citealt{Nadler191203303,Patel200101746,Vasiliev230604837}). This result is also broadly consistent with previous predictions for the number of LMC-associated satellites (e.g., \citealt{Deason150404372,Dooley170305321,Jahn190702979}).

Using two MW-like simulations with LMC analogs from the original \cite{MWW2015} Symphony MW suite, \cite{Nadler191203303} demonstrated that spatial anisotropy of subhalo and satellite galaxy populations driven by the LMC system is necessary to fit the observed MW satellite population. The LMC analogs studied in \cite{Nadler191203303} have sub-to-host halo mass ratios of $0.1$ and $0.2$, respectively, so they enhance the SHMF, inner radial distribution, and spatial anisotropy by a typical amount relative to our LMC analog sample (see Figure~\ref{fig:frac_associated}). Our LMC-associated subhalo results are also consistent with \cite{Lu160502075}, who found a similar enhancement of the satellite stellar mass function and spatial anisotropy due to satellites accreted with LMC analogs in the original Symphony MW suite \citep{MWW2015}. 

We note that \cite{Barry230305527} reports even larger LMC-related SHMF enhancements using hydrodynamic zoom-in simulations run with the Feedback in Realistic Environments (FIRE) code. The central galaxies in these simulations preferentially disrupt subhalos that accrete earlier \citep{Garrison-Kimmel170103792,Nadler171204467}, which may enhance the relative contribution to the SHMF from LMC-associated subhalos. Future work will explore this effect by resimulating Symphony MW and Milky Way-est hosts with embedded disk potentials (Y. Wang et al. 2024, in preparation).

\subsubsection{Observational Implications}
\label{sec:obs_implications}

Accurate subhalo population predictions are crucial for analyses of the MW's satellite galaxy and stellar stream populations (e.g., \citealt{Drlica-Wagner190201055,Banerjee220307049}). We have analyzed subhalos with present-day virial masses above $\approx 10^8~\msun$, most of which are expected to host satellite galaxies (e.g., \citealt{Nadler191203303,Ahvazi230813599}). 
Our predictions are therefore relevant for interpreting upcoming MW satellite observations, particularly in the southern hemisphere, where most LMC-associated satellites are located (e.g., \citealt{Jethwa160304420,Drlica-Wagner191203302,Drlica-Wagner220316565}). Intriguingly, the centrally concentrated nature of Milky Way-est subhalo populations may help explain the large number of satellite galaxies observed in the MW's inner regions (e.g., \citealt{Yniguez13050560,Graus180803654}). It will be interesting to revisit galaxy--halo connection studies and dark matter constraints from satellite galaxies using the Milky Way-est suite.

Our predictions are also relevant for interpreting satellite galaxy populations beyond the MW. In particular, recent surveys have characterized satellite populations of MW-mass centrals throughout the Local Volume (e.g., ELVES; \citealt{Carlsten220300014}) and at distances of $25$--$40~\mpc$ (e.g., SAGA; \citealt{Geha170506743,Mao200812783,Mao240414498}). Interestingly, SAGA satellite abundances correlate strongly with the brightest satellite luminosity, which is reminiscent of the strong impact that Milky Way-est LMC analogs have on total subhalo abundance. Thus, we expect that Milky Way-est simulations will inform future comparisons between MW, Local Volume, and SAGA satellite populations.

Beyond satellite galaxy modeling, we anticipate that Milky Way-est simulations will help refine predictions for the MW's stellar stream population, particularly for the subhalos that perturb stellar streams. Current stream observations are most sensitive to subhalos in the inner regions of the MW (e.g., \citealt{Bonaca181103631,Banik191102662}). Thus, the abundant and centrally concentrated nature of Milky Way-est subhalo populations relative to typical MW-mass hosts has important implications for stream modeling. Spatial anisotropy in the distribution of stream perturbers over the full sky may also become detectable as stream observations improve \citep{Drlica-Wagner190201055}. Resimulations of Milky Way-est systems at higher resolution will resolve even lower-mass subhalos associated with the LMC, which will be relevant for interpreting upcoming stellar stream measurements (e.g., \citealt{Arora230915998}).

\subsection{Simultaneously Modeling the Impact of GSE and LMC}
\label{sec:unusual}

A key strength of our work is that both GSE and LMC analogs are explicitly included in Milky Way-est halo selection criteria. Although we mainly focused on GSE and LMC's individual contributions to Milky Way-est hosts' formation histories and subhalo populations, several correlated effects may merit dedicated study in future work.

First, Milky Way-est hosts' specific dark matter accretion rates are enhanced during the GSE merger, consistent with previous simulation results (e.g., \citealt{Evans200504969}). However, during this epoch, Milky Way-est hosts' masses are lower than average hosts with the same final mass due to the LMC's large contribution to Milky Way-est hosts' total masses at late times. For example, the mean mass of Milky Way-est hosts is $1.0\times 10^{11}~\msun$ at $a=0.25$ ($z=3$), versus $1.6\times 10^{11}~\msun$ in Symphony MW. Thus, it is timely to revisit the impact of GSE on disk and stellar halo formation in low-mass MW progenitors.

Second, various studies have shown that only a small fraction ($\approx 10\%$) of Milky Way-mass halos in cosmological simulations host LMC- and SMC-mass subhalos (e.g., \citealt{BoylanKolchin09114484,Busha10116373,Liu10112255}). We find that an even smaller fraction ($<1\%$) of hosts have both recently accreted, nearby LMCs and GSE analogs, where the LMC's present-day distance is the most constraining cut (see Table~\ref{tab:1}). Qualitatively, this result is not surprising: adding constraints makes the resulting sample rarer. Practically, the number density of hosts that satisfy our criteria is $33$ systems in a $(125~\mpc~h^{-1})^3$ volume. Thus, simulating a large number of halos with both GSE and LMC analogs at sufficient resolution to study their low-mass subhalo populations is challenging. Future work directly generating MW-like hosts with LMC and GSE analogs and perhaps additional systems like SMC analogs will be useful. Constrained simulations (e.g., \citealt{Rey190904664}) and semi-analytic merger-tree realizations (e.g., \citealt{Nadler221208584}) are promising in this regard.

\subsection{Caveats}
\label{sec:future_work}

We now discuss several caveats. First, our simulations do not include baryons, which are known to affect subhalo populations, e.g.\ via enhanced disruption of subhalos that orbit near the central galaxy (e.g., \citealt{DOnghia08020001,Garrison-Kimmel170103792,Kelley181112413}). 
Unlike existing zoom-in simulations where this disk potential technique has been applied, Milky Way-est systems are explicitly selected to accrete both GSE and LMC analogs. Incorporating disk potentials in resimulations of Milky Way-est systems will be a step toward modeling the impact of baryons on their subhalo populations. Because the disk preferentially disrupts subhalos that accrete early \citep{Garrison-Kimmel170103792,Nadler171204467}, including central galaxy potentials will likely enhance the contribution of recently accreted LMC-associated subhalos to host subhalo populations, particularly in the inner regions, while further suppressing the contribution of GSE-associated subhalos.

Second, we have defined ``MW-like'' halos in our Milky Way-est suite based on halo mass, concentration, and the existence of GSE and LMC analogs. Even with all of these criteria, other features of Milky Way-est systems' accretion histories may not be representative of the MW. For example, as noted in Section~\ref{sec:sample}, we have not required Milky Way-est hosts to accrete a Sgr analog or remain quiescent between GSE disruption and the accretion of a potential Sgr analog; for Milky Way-est hosts with potential Sgr analogs, it will be interesting to explore how many Sgr-associated subhalos survive as the system tidally disrupts. Meanwhile, large-scale environment---which is not directly captured by our LMC and GSE constraints---may further influence host halo formation histories (e.g., \citealt{Garrison-Kimmel190310515,Santistevan200103178}). Beyond dark matter assembly histories, it is unclear whether Milky Way-est hosts form galaxies consistent with the MW galaxy's observed properties. We plan to apply empirical (e.g., \textsc{UniverseMachine}; \citealt{Behroozi180607893,Wang210211876,Wang240414500}) and semi-analytic (e.g., \textsc{GRUMPY}; \citealt{Kravtsov210609724}) models to gain insight into galaxy formation in Milky Way-est systems. Milky Way-est systems are excellent candidates for zoom-in resimulation with hydrodynamic and galaxy formation physics.

Finally, our analysis may underestimate the contribution of GSE-associated subhalos. In particular, although \textsc{Rockstar} and \textsc{consistent-trees} perform well compared to other subhalo finders and merger-tree algorithms (e.g., \citealt{Onions12033695,Srisawat13073577}), these tools can lose track of early-accreted subhalos in a nonphysical manner \citep{Mansfield230810926}. It will be important to study GSE analogs and their contribution to present-day Milky Way-est subhalo populations using improved halo catalogs and merger trees based on, for example, particle tracking.


\section{Conclusions}
\label{sec:conclusion}

We have presented Milky Way-est, a suite of 20 cosmological cold-dark-matter-only zoom-in simulations of Milky Way analogs selected to undergo LMC- and GSE-like merger events. 
By comparing Milky Way-est with hosts that are not specifically selected to undergo these mergers, both from their parent cosmological volume and from the Symphony MW zoom-in simulations \citep{Nadler_2023}, we find the following: 
\begin{itemize}
    \item Given our selection criteria, host halos with both LMC and GSE analogs constitute $<1\%$ of all MW-mass systems (i.e., $33$ out of $4455$ MW-mass systems in a $(125~\mpc~h^{-1})^3$ volume; see Table~\ref{tab:1}).
    \item Compared to Symphony MW, Milky Way-est systems' mean half-mass formation times are shifted later, primarily due to LMC accretion (Figure~\ref{fig:mahs}).
    \item Milky Way-est systems host $22\%$ more subhalos down to a fixed sub-to-host halo mass ratio compared to Symphony MW hosts, which are not explicitly selected to satisfy MW-like criteria beyond host halo mass; this enhancement reaches $\approx 80\%$ in the inner $100~\kpc$ (Figure~\ref{fig:shmf_raddistri}). 
    \item These SHMF and radial distribution differences can largely, if not entirely, be explained by LMC-associated subhalos in Milky Way-est, while GSE-associated subhalos do not contribute significantly to present-day Milky Way-est subhalo populations.
    \item Milky Way-est systems have spatially anisotropic subhalo populations, with $\approx 60\%$ of the total subhalo population in the inner $100~\kpc$ found in the current direction of the LMC (Figure~\ref{fig:anisotropy}).
\end{itemize}

Our LMC analogs are the main drivers of these formation history and subhalo population effects. In particular, LMC-associated subhalos enhance the abundance, inner radial distribution, and spatial anisotropy of Milky Way-est subhalo populations relative to typical MW-mass hosts, building on previous results (e.g., \citealt{Sales11050425,Santos-Santos201113500,Nadler191203303,Nadler210912120,Barry230305527}). Furthermore, our results suggest that the LMC's orbit correlates with its impact on the MW subhalo population.

Milky Way-est GSE analogs do not substantially affect the host halo formation history and subhalo population statistics we have studied. Nonetheless, we expect that Milky Way-est simulations will provide insights into GSE's contribution to ancient substructure in the inner regions of the MW, the formation of the MW halo, and the MW's dark matter phase-space distribution, following recent studies (e.g., \citealt{Bose190904039,Fattahi181007779,Evans200504969,Dillamore210913244}).

It is interesting to consider how many of (and how accurately) the MW's major accretion events must be modeled to meet the precision of upcoming MW dwarf galaxy and stellar stream observations.  
On even smaller, subkiloparsec scales, stellar velocity distributions near the solar circle and dark matter direct-detection experiments are sensitive to substructure accreted during specific merger and accretion events like the GSE and LMC (e.g., \citealt{Necib181012301,Besla190104140,Smith-Orlik230204281}). This motivates studies of the local dark matter distribution using Milky Way-est.


\section*{Acknowledgements}

Milky Way-est data products are publicly available at \url{http://web.stanford.edu/group/gfc/gfcsims/}.

We thank the anonymous referee for their valuable feedback. We thank Matthew Becker for his work on the Chinchilla simulation suite used in this study. We are grateful to Elise Darragh-Ford, Nitya Kallivayalil, Philip Mansfield, Ekta Patel, Yunchong Wang, and Andrew Wetzel for comments on the manuscript and to Susmita Adhikari, Christian Aganze, and Arka Banerjee for helpful conversations related to this work.

This research was supported in part by the Kavli Institute for Particle Astrophysics and Cosmology (KIPAC) at Stanford University and SLAC National Accelerator Laboratory and by the U.S.\ Department of Energy under contract number DE-AC02-76SF00515 to SLAC National Accelerator Laboratory. D.B.\ received additional support from the Stanford University Summer Undergraduate Research Program in Physics. This work was performed in part at the Aspen Center for Physics, which is supported by National Science Foundation grant No. PHY-2210452.

This research made use of computational resources at SLAC National Accelerator Laboratory, a U.S.\ Department of Energy Office of Science laboratory, and the Sherlock cluster at the Stanford Research Computing Center (SRCC); the authors are thankful for the support of the SLAC and SRCC computational teams. This work used the Extreme Science and Engineering Discovery Environment (XSEDE) Stampede2 and Ranch systems at the Texas Advanced Computing Center (TACC) through allocation TG-PHY200009. The authors acknowledge TACC at The University of Texas at Austin for providing HPC resources that have contributed to the research results reported within this paper.

This work used data from the Symphony suite of simulations (\url{http://web.stanford.edu/group/gfc/symphony/}), which was supported by the Kavli Institute for Particle Astrophysics and Cosmology at Stanford University and SLAC National Accelerator Laboratory, and by the U.S.\ Department of Energy under contract number DE-AC02-76SF00515 to SLAC National Accelerator Laboratory.

This research used \url{https://arXiv.org} and NASA's Astrophysics Data System for bibliographic information.

\bibliographystyle{yahapj2}
\bibliography{references}


\begin{deluxetable*}{{c@{\hspace{0.07in}}c@{\hspace{0.07in}}c@{\hspace{0.07in}}c@{\hspace{0.07in}}c@{\hspace{0.07in}}c@{\hspace{0.07in}}c@{\hspace{0.07in}}c@{\hspace{0.07in}}c@{\hspace{0.07in}}c@{\hspace{0.07in}}c@{\hspace{0.07in}}c@{\hspace{0.07in}}c@{\hspace{0.07in}}c}}[t!]
\centering
\tablecolumns{14}
\tablecaption{Properties of Milky Way-est Simulations.}
\tablehead{
\colhead{Name}
& \colhead{$a_{\mathrm{LMC,50}}$}
& \colhead{$\frac{M_{\mathrm{host}}}{10^{12}~\msun}$}
& \colhead{$\frac{R_{\mathrm{vir,host}}}{\kpc}$}
& \colhead{$c_{\mathrm{host}}$}
& \colhead{$a_{\mathrm{1/2,host}}$}
& \colhead{$\frac{M_{\mathrm{LMC}}}{10^{11}~\msun}$}
& \colhead{$\frac{M_{\mathrm{peak,LMC}}}{10^{11}~\msun}$}
& \colhead{$\frac{r_{\mathrm{LMC}}}{\kpc}$}
& \colhead{$a_{\mathrm{disrupt,GSE}}$} 
& \colhead{$\frac{M_{\mathrm{peak,GSE}}}{10^{11}~\msun}$}
& \colhead{$N_{\mathrm{sub,tot}}$}
& \colhead{$N_{\mathrm{sub,LMC}}$}
& \colhead{$N_{\mathrm{sub,GSE}}$}}
\startdata
Halo169
& 0.97
& 0.98
& 255
& 8.2
& 0.62
& 1.85
& 2.29
& 58
& 0.44
& 0.94
& 125
& 20
& 4\\
Halo453
& 0.96
& 1.00
& 255
& 12.7
& 0.67
& 2.99
& 3.30
& 57
& 0.31
& 0.54
& 79
& 13
& 0\\
Halo004
& 1.00
& 1.03
& 263
& 11.2
& 0.52
& 1.50
& 1.74
& 61
& 0.41
& 1.4
& 94
& 10
& 4\\
Halo476
& 0.96
& 1.06
& 260
& 12.5
& 0.49
& 2.04
& 2.50
& 50
& 0.5
& 1.1
& 109
& 13
& 5\\
Halo113
& 1.01
& 1.08
& 269
& 11.6
& 0.50
& 0.28
& 0.36
& 49
& 0.40
& 1.1
& 75
& 0
& 1\\
Halo222
& 1.05
& 1.12
& 278
& 10.8
& 0.63
& 2.25
& 2.67
& 63
& 0.38
& 1.2
& 117
& 15
& 1\\
Halo407
& 1.00
& 1.13
& 271
& 8.4
& 0.45
& 0.78
& 0.96
& 54
& 0.46
& 0.5
& 110
& 2
& 1\\
Halo327
& 0.94
& 1.13
& 261
& 11.4
& 0.50
& 1.16
& 1.41
& 44
& 0.5
& 2.4
& 69
& 6
& 8\\
Halo975
& 0.99
& 1.14
& 270
& 13.9
& 0.72
& 2.91
& 3.15
& 34
& 0.32
& 0.53
& 81
& 16
& 1\\
Halo170
& 0.97
& 1.21
& 273
& 12.1
& 0.59
& 2.16
& 2.61
& 50
& 0.29
& 0.4
& 101
& 19
& 0\\
Halo719
& 1.01
& 1.23
& 281
& 9.7
& 0.86
& 3.98
& 4.63
& 51
& 0.39
& 0.88
& 127
& 17
& 0\\
Halo282
& 1.04
& 1.28
& 289
& 9.0
& 0.52
& 0.53
& 0.71
& 38
& 0.44
& 0.95
& 117
& 3
& 2\\
Halo983
& 1.00
& 1.37
& 289
& 10.1
& 0.67
& 1.80
& 2.50
& 49
& 0.33
& 0.53
& 105
& 8
& 3\\
Halo349
& 0.90
& 1.40
& 274
& 11.7
& 0.51
& 2.33
& 2.90
& 52
& 0.48
& 0.15
& 119 
& 16
& 0\\
Halo747
& 0.97
& 1.41
& 288
& 9.6
& 0.50
& 0.59
& 0.73
& 54
& 0.47
& 1.3
& 132
& 4
& 4\\
Halo666
& 1.00
& 1.55
& 302
& 8.0
& 0.69
& 6.12
& 6.36
& 79
& 0.42
& 1.7
& 154
& 34
& 9\\
Halo659
& 0.96
& 1.56
& 296
& 10.5
& 0.54
& 0.36
& 0.80
& 44
& 0.34
& 0.69
& 136
& 2
& 1\\
Halo788
& 1.00
& 1.71
& 311
& 11.9
& 0.49
& 0.49
& 0.68
& 34
& 0.33
& 1.0
& 162
& 5
& 0\\
Halo756
& 1.00
& 1.73
& 313
& 10.5
& 0.43
& 1.34
& 1.61
& 66
& 0.45
& 3.4
& 147
& 15
& 5\\
Halo229
& 1.01
& 1.74
& 316
& 12.3
& 0.48
& 0.10
& 0.19
& 64
& 0.41
& 0.6
& 143
& 1\
& 1\
\enddata
{\footnotesize \tablecomments{The columns, from left to right, list the name of each Milky Way-est host; the scale factor when each simulation is analyzed; the MW host halo virial mass, host virial radius, host concentration, and half-mass scale factor; the LMC analog present-day and peak virial mass, and distance; the GSE analog disruption scale factor and peak mass; and the number of total, LMC-associated, and GSE-associated subhalos with $M_{\mathrm{sub}}>1.2\times 10^8~\msun$ (where LMC/GSE-associated subhalos are identified within the LMC/GSE virial radius at the time step immediately before LMC/GSE accretion into the MW virial radius). All quantities except $M_{\mathrm{peak,LMC}}$ and $M_{\mathrm{peak,GSE}}$ are evaluated at (or relative to) $a_{\mathrm{LMC,50}}$. The table is sorted in order of increasing $M_\mathrm{host}$.}}
\label{tab:sims}
\end{deluxetable*}

\appendix

\section{Host Properties}
\label{sec:host_properties}

Table~\ref{tab:sims} lists host halo, LMC analog, and GSE analog properties for all $20$ Milky Way-est hosts.

\section{Convergence Test}
\label{sec:convergence}

\begin{figure}[t!]
\includegraphics[trim={0 0cm 0 0},width=0.49\textwidth]{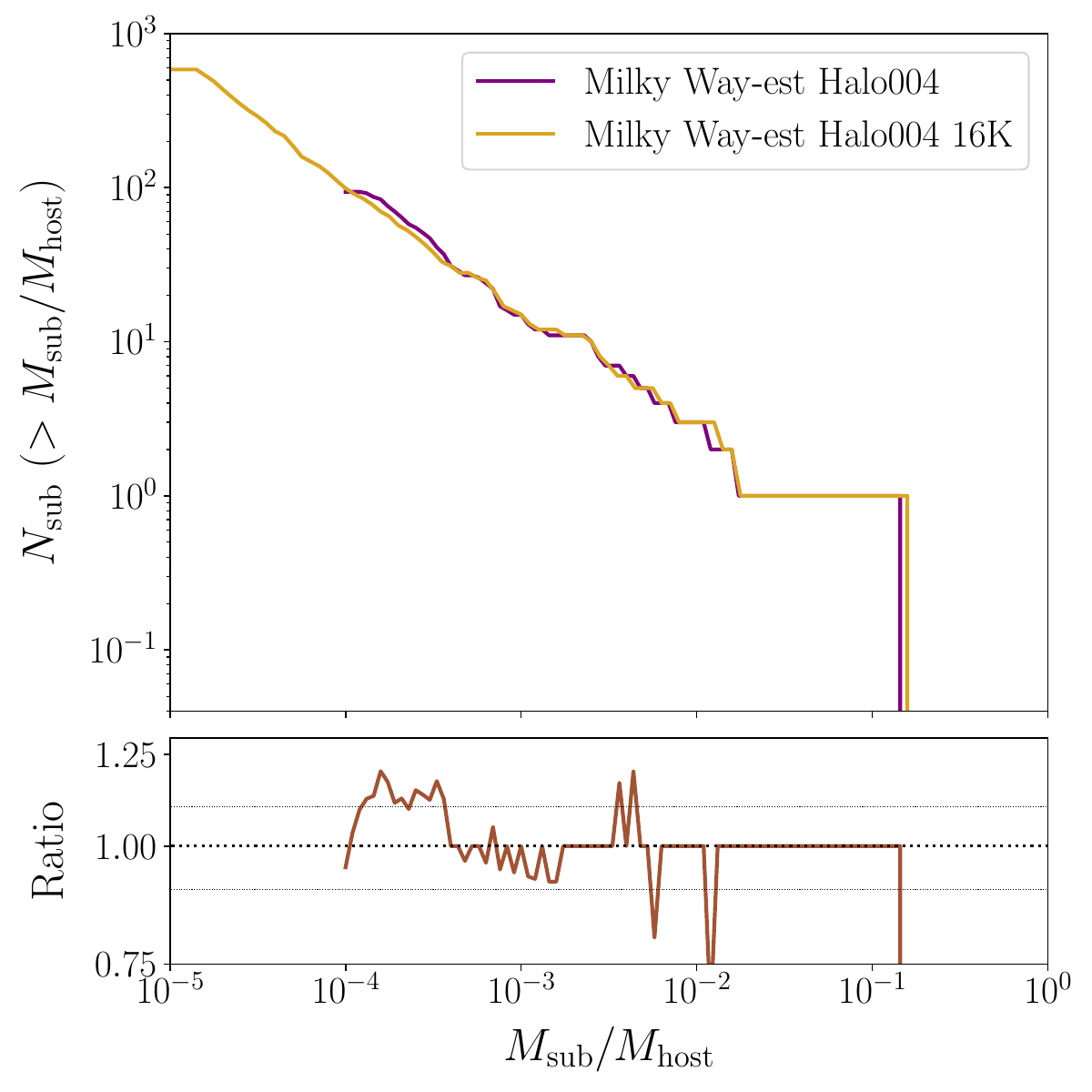}
\caption{SHMF for Milky Way-est Halo004 at our fiducial resolution (purple) compared to a higher-resolution resimulation (gold). For the fiducial-resolution case, we use our standard subhalo mass cut of $M_{\mathrm{sub}}>300\times m_{\mathrm{part}}=1.2\times10^8~\msun$; for the higher-resolution case, we use $M_{\mathrm{sub}}>300\times m_{\mathrm{part,high-res}}=1.5\times10^7~\msun$. The bottom panel shows the ratio of fiducial to higher-resolution measurements, along with $\pm 10\%$ deviations (thin dotted lines) about a ratio of 1 (dotted line).}
\label{fig:16k}
\end{figure}

Here, we perform a higher-resolution resimulation of one Milky Way-est system to test for convergence. In particular, we generate high-resolution initial conditions for Halo004 with an equivalent of $16,384$ particles per side in the most refined region, with a corresponding dark matter particle mass of $m_{\mathrm{part,high-res}}=5.0\times 10^4~ \msun$. We run this higher-resolution simulation with a comoving Plummer-equivalent gravitational softening of $80~\pc~h^{-1}$.

As expected, we find that the high-resolution host halo's MAH matches our fiducial-resolution simulation result extremely well at late times. The MAHs match reasonably well even at early times; both the high- and fiducial-resolution MW progenitors are resolved with $>300$ particles up to $z \approx 15$. 
SHMFs are consistent among resolution levels within $\approx 10\%$ down to our fiducial-resolution $M_{\mathrm{sub}}$ threshold, consistent with previous convergence tests for Symphony (\citealt{Nadler_2023}; see Figure~\ref{fig:16k}).

We also compare radial distributions between resolution levels, finding that they are converged for $r\gtrsim0.4R_{\mathrm{vir,host}}$; at smaller radii, our fiducial-resolution simulation's radial distribution is more concentrated than the higher-resolution result, when compared down to a fixed mass threshold. We caution that this radial distribution comparison is difficult to interpret for a single halo because subhalos' orbital phases can change between resolution levels. Thus, we leave a detailed resolution study using a larger sample of hosts to future work.

\end{document}